%% file: quantumcomp.tex
\documentclass[conference]{IEEEtran}

\IEEEoverridecommandlockouts
\usepackage{cite}
\usepackage{amsmath,amssymb,amsfonts}
\usepackage{algorithmic}
\usepackage{graphicx}
\usepackage{textcomp}
\usepackage[table]{xcolor}
\usepackage{booktabs}
\usepackage{tabularx}
\usepackage{import}
\usepackage{booktabs}
\usepackage{soul}
\usepackage{tcolorbox}
\colorlet{Mycolor1}{green!10!orange!90!}
\colorlet{Mycolor2}{green!10!orange!90!}

\def\BibTeX{{\rm B\kern-.05em{\sc i\kern-.025em b}\kern-.08em
    T\kern-.1667em\lower.7ex\hbox{E}\kern-.125emX}}
    
\begin{document}

\newcommand{\heng}[1]{\textcolor{red}{{\it [Heng says: #1]}}}
\newcommand{\Foutse}[1]{\textcolor{blue}{{\it [Foutse says: #1]}}}
\newcommand{\raed}[1]{\textcolor{Mycolor1}{{\it #1}}}
\newcommand{\fix}[1]{\textcolor{Mycolor2}{{\it #1}}}
%\newcommand{\raed}[1]{}

%\title{Understanding Quantum Software Engineering Challenges: An Empirical Study of Technical Q\&As and GitHub Issues}
\title{\huge{Understanding Quantum Software Engineering Challenges}\\ \Large{An Empirical Study on Stack Exchange Forums and GitHub Issues}}
\author{
\IEEEauthorblockN{Mohamed Raed El aoun, Heng Li, Foutse Khomh, Moses Openja %Given Name Surname
}
\IEEEauthorblockA{\textit{Department of Computer Engineering and Software Engineering} \\
\textit{Polytechnique Montréal, Montréal, QC, Canada}\\
\{mohamed-raed.el-aoun, heng.li, foutse.khomh, moses.openja\}@polymtl.ca}}

\begin{comment}
%1\textsuperscript{st} 
\author{\IEEEauthorblockN{Mohamed Raed El aoun %Given Name Surname
}
\IEEEauthorblockA{\textit{Dept. of Software Engineering} \\
\textit{Polytechnique Montréal, Canada}\\
mohamed-raed.el-aoun@polymtl.ca}
\and
\IEEEauthorblockN{Heng Li %Given Name Surname
}
\IEEEauthorblockA{\textit{Dept. of Software Engineering} \\
\textit{Polytechnique Montréal, Canada}\\
heng.li@polymtl.ca}
\and
\IEEEauthorblockN{Foutse Khomh %Given Name Surname
}
\IEEEauthorblockA{\textit{Dept. of Software Engineering} \\
\textit{Polytechnique Montréal, Canada}\\
foutse.khomh@polymtl.ca}
}
\end{comment}

\maketitle

\begin{abstract}
With the advance of quantum computing, quantum software becomes critical for exploring the full potential of quantum computing systems.
Recently, quantum software engineering (QSE) becomes an emerging area attracting more and more attention. However, it is not clear what are the challenges and opportunities of quantum computing facing the software engineering community. 
%Therefore, in this work, we make an initial effort to empirically study the existing practices of quantum software engineering.
This work aims to understand the QSE-related challenges perceived by developers.
We perform an empirical study on Stack Exchange forums where developers post-QSE-related questions \& answers and Github issue reports where developers raise QSE-related issues in practical quantum computing projects. 
Based on an existing taxonomy of question types on Stack Overflow, we first perform a qualitative analysis of the types of QSE-related questions asked on Stack Exchange forums. 
We then use automated topic modeling to uncover the topics in QSE-related Stack Exchange posts and GitHub issue reports. 
Our study highlights some particularly challenging areas of QSE that are different from that of traditional software engineering, such as explaining the theory behind quantum computing code, interpreting quantum program outputs, and bridging the knowledge gap between quantum computing and classical computing, as well as their associated opportunities.

\end{abstract}


\input{Sections/introduction}

\input{Sections/background}

\input{Sections/experimentSetup}

\input{Sections/result}

\input{Sections/Threats}

\input{Sections/Conclusions}

\bibliographystyle{IEEEtran}
%\bibliography{IEEEabrv,mybibfile}
\bibliography{quantumcomp}

\end{document}

%% file: Sections/introduction.tex
\vspace{-5pt}
\section{Introduction}

Over the past decades, quantum computing has made steady and remarkable progress~\cite{knight2018serious, maslov2018outlook, zhao2020quantum}. For example, IBM Quantum\cite{ibm} now supports developers to develop quantum applications using its programming framework and execute them on its cloud-based quantum computers.
Based on the quantum mechanics principles of \texttt{superposition} (quantum objects can be in different states at the same time)~\cite{dirac1981principles} and \texttt{entanglement} (quantum objects can be deeply connected without direct physical interaction)~\cite{schrodinger1935discussion}, quantum computers are expected to make revolutionary computation improvement over today's classical computers~\cite{mueck2017quantum}. In particular, quantum computing is expected to help solve the computational problems that are difficult for today's classical computers, including problems in cryptography, chemistry, financial services, medicine, and national security~\cite{piattini2020talavera}. 

The success of quantum computing will not be accomplished without quantum software. Several quantum programming languages (e.g., QCL~\cite{omer2003qcl}) and development tools (e.g., Qiskit\cite{Qiskit} have been developed since the first quantum computers. 
Large software companies like Google\cite{quantumai}, IBM\cite{ibm}, and Microsoft\cite{microsoft.azure} have developed their technologies for quantum software development.
Quantum software developers have also achieved some preliminary success in applying quantum software to certain computational areas (e.g, machine learning~\cite{biamonte2017quantum}, optimization~\cite{guerreschi2017practical}, cryptography~\cite{mailloux2016post}, and chemistry~\cite{reiher2017elucidating}).
However, there still lacks large-scale quantum software.
Much like Software Engineering is needed for developing large-scale traditional software, the concept of Quantum Software Engineering (QSE) has been proposed to support and guide the development of large-scale, industrial-level quantum software applications. This concept has been gaining %The concept of quantum software engineering (QSE) has been proposed to advance development of quantum software applications and start to gain 
more and more attention recently~\cite{piattini2020talavera,zhao2020quantum, piattini2020quantum}. QSE aims to apply or adapt existing software engineering processes, methods, techniques, practices, and principles to the development of quantum software applications, or create new ones~\cite{piattini2020talavera}.
Pioneering work sheds light on new directions for QSE, such as quantum software processes \& methodologies~\cite{moguel2020roadmap}, quantum software modeling~\cite{barbosa2020software}, and design of quantum hybrid systems~\cite{piattini2021toward}. 
In the meanwhile, we observe an exponential increase of discussions related to quantum software development on technical Q\&A forums such as Stack Overflow(e.g. from 8 in 2010 to 1434 in 2020). 
We also notice an increasing number of quantum software projects hosted on GitHub, where developers use issue reports to track their development and issue fixing processes. 
Such technical Q\&As and issue reports may communicate developers' faced challenges when developing quantum software applications. 

In this paper, we aim to understand the challenges perceived by quantum software developers and seek opportunities for future QSE research and practice. In particular, we examine technical Q\&A forums where developers ask QSE-related questions, and GitHub issue reports where developers raise QSE-related issues. 
%We extract Q\&A posts from four Stack Exchange forums \heng{all our forums are Stack Exchange forums, right?} and issue reports from 122 quantum software projects. 
We apply a series of heuristics to search and filter Q\&A posts that are related to QSE and to search and filter GitHub projects that are related to quantum software.
In total, we extract and analyze 3,117 Q\&A posts and 43,979 Github issues that are related to QSE. We combine manual analysis and automated topic modeling to examine these Q\&A posts and Github issues, to understand the QSE challenges developers are facing.
%Furthermore,we will not only discuss the topics in QSE but also we are considering to analyse the reasons the users are asking the questions by categorizing the questions based on why or How in order to find the role that the posts have in QSE paradigm.
%For that purpose
In particular, our study aims to answer the three following research questions (RQs):
%\newline 

\begin{description}

\item [\textbf{RQ1:}] \textit{What types of QSE questions are asked on technical forums?}

To understand the intention behind developers' questions on technical forums and the types of information that they are seeking, we manually examined a statistically representative sample of questions.
We extended a previous taxonomy from prior work \cite{Beyer2021WhatKO} and found nine categories of questions.
Our results highlight the need for future efforts to support developers' quantum program development, in particular, to develop learning resources,  to help developers fix errors, and to explain the theory behind quantum computing code.

\item [\textbf{RQ2:}] \textit{What QSE topics are raised in technical forums?} 
%In order to understand the challenges that developers are facing when developing quantum software applications
%We use topic modeling on SO and SE posts and answers to identify the discussed topics related to QSE. We end up with 9 LDA-topics where we observe \texttt{environment management}, \texttt{Dependency management} and \texttt{Algorithm complexity} are the most discussed topics. Moreover we study the topics popularity among the developer and how difficult to get an accepted answer. As a result we find out \texttt{quantum vs. classical computing} is the most popular topic while \texttt{environment management} is the most difficult topic 
The QSE-related posts may reflect developers' challenges when learning or developing quantum programs. To understand their faced challenges, we use topic models to extract the semantic topics in their posts. We derived nine topics including traditional software engineering topics (e.g., \texttt{environment management} and \texttt{dependency management}) and QSE-specific topics (e.g., \texttt{quantum execution results} and \texttt{quantum vs. classical computing}).
We highlighted some particularly challenging areas for QSE, such as interpreting quantum program outputs, understanding quantum algorithm complexity, and bridging the knowledge gap between quantum computing and classical computing.

%\newline
\item [\textbf{RQ3:}] \textit{What QSE topics are raised in the issue reports of quantum-computing projects?} 
Issue reports of quantum computing projects record developers' concerns and discussions when developing these projects. Thus, we analyze the topics in the issue reports to understand the challenges are developers facing in practical quantum computing projects. % as well as the prevalence of these challenges.
We observe that the QSE-related challenges that we derived from forum posts indeed impact practical quantum program development in these GitHub projects, while GitHub issues bring new perspectives on developers' faced challenges (e.g., on specific quantum computing applications such as machine learning).
We also observe that such challenges are general among quantum computing projects. 
%The textual information in the issue reports may communicate developers' challenges when developing quantum computing applications. 

\end{description}

\noindent \textbf{Paper organization.} The rest of the paper is organized as follows. In Section~\ref{sec:background} we discuss the background about quantum software engineering and the related work. Then, in Section~\ref{sec:setup} we describe the design of our study. %our experiments for collecting and analyzing the data. 
In Section~\ref{sec:results} we present our results. %for answering our research questions. 
Section~\ref{sec:threats} discusses threats to the validity of our findings. Finally, Section~\ref{sec:conclusions} concludes the paper.

%% file: Sections/background.tex
\section{Background and Related Work} \label{sec:background}
%The main purpose of this work is to investigate the topics discussed around quantum computing and software engineering and find out what are the challenges this paradigm is facing. 
This study aims to understand the quantum software engineering challenges through examining technical forum posts and GitHub issue reports. %\Foutse{this is too strong...we dont really study the practices...we examine the challenges instead}
In this section, we present the background and prior work related to our study. 
First, we describe the background and related work of quantum computing, quantum programming, and quantum software engineering.
Then, we discuss prior work that performs topic analysis on technical forum posts and issue reports.

\subsection{Quantum Computing}
%\heng{to revisit}
%\heng{Always need space between sentences!}

%Quantum computing is a computing paradigm exploiting \Foutse{i think it builds on two core principle: entanglement and superposition...please revise the definition to make it precise!}
%the ability to harness most of the phenomena of quantum mechanics to provide faster processing powers. The devices that execute the quantum theories \Foutse{theories?} 
%are called quantum computers\heng{sometimes you say quantum computers, sometimes quantum systems, and sometimes quantum machines. Just use a consistent term}.
Quantum computers aim to leverage the principles of quantum mechanics such as \texttt{superposition} and \texttt{entanglement} to provide computing speed faster than today's classical computers.
While classical computers use bits in the form of electrical pulses to represent 1s and 0s, quantum computers use quantum bits or \textbf{Qubits} in the form of subatomic particles such as electrons or photons to represent 1s and 0s. 
A Qubit, unlike a classical bit, can be 0 or 1 with a certain probability, which is known as the \textbf{superposition} principle~\cite{kaye2007introduction}.
%In other words, a Qubit can represents multiple status of 0s and 1s at the same time.
In other words, a quantum computer consisting of Qubits is in many different states at the same time.
When a Qubit is \textbf{measured}, it collapses into a deterministic classical state.
%the ability of quantum computers to put Qubits simultaneously in multiple states is known as superposition \cite{kaye2007introduction}.
%\heng{needs explanation and/or reference for what is superposition}. 
The status of two or more of Qubits can be correlated (or entangled) in the sense that changing the status of one Qubit will change the status of the others in a predictable way, which is known as the \textbf{entanglement} phenomenon~\cite{kaye2007introduction}.  
%Therefore, a computer with a big amount of Qubits in multiple states can perform parallel computation in a large scale.
The \texttt{superposition} and \texttt{entanglement} phenomenons give quantum computers advantages over classical computers in performing large-scale parallel computation~\cite{kaye2007introduction}.

%One of the most powerful property of quantum computing is entanglement. Qubits are called entangled when a pair of Qubits exist in a single quantum state and changing the state in one pair will change the state in the other pair in a predictable way \cite{kaye2007introduction}. %\heng{My understanding: A pair of Qubits are called entangled Qubits when changing the state of one Qbit will change the state of the other in a predictable way.} 
%Thanks to this property quantum machine have shown an exponential increase in their abilities.
%\heng{$\leftarrow$ not clear why, needs a brief explanation, is it because there are many possible states?}.

%The quantum systems with high number of entangled Qubits are impossible to models with classical computers%\heng{$\leftarrow$needs explanation and/or reference}. 
%However to reach high capabilities than classical computer it's mandatory to achieve big amounts of entanglement%\heng{$\leftarrow$needs explanation and/or reference}. 
%Therefore Quantum algorithm must exploit this a huge amount of this property.
%\heng{$\leftarrow$not clear what it means}.

\begin{figure}[htbp]
\vspace{-3mm}
\centerline{\includegraphics[width=0.49\textwidth]{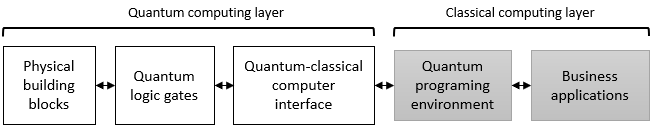}}
\vspace{-5pt}
\caption{The architecture of a quantum computer \cite{DBLP:journals/corr/abs-1803-07407,zhao2020quantum}}.
\label{fig:quantumcomp}
\vspace{-8mm}
\end{figure}

Similar to classical logic gates (e.g., \texttt{AND}, \texttt{OR}, \texttt{NOT}), \textbf{quantum logic gates} (or \textbf{quantum gates}) alter the states (the probability of being 0 or 1) of the input Qubits.
Like classical digit circuits, \textbf{quantum circuits} are collections of quantum logic gates interconnected by quantum wires. 
%\heng{too high level. see comments in the figure title}
Figure~\ref{fig:quantumcomp} illustrates the architecture of a quantum computer \cite{DBLP:journals/corr/abs-1803-07407,zhao2020quantum}. The architecture contains two layers: a quantum computing layer where the quantum physics and circuits reside, and a classical computing layer where the quantum programming environment and software applications reside.
\begin{itemize}
    \item \emph{Physical building blocks}: physical realization of Qubits and their coupling/interconnect circuitry. %the hardware responsible for physical realization of Qubits in addition to the physical Qubits coupler which are required for control operation and Qubit addressing \cite{DBLP:journals/corr/abs-1803-07407}.%\heng{punctuation here and below}
    \item \emph{Quantum logic gates}: physical circuitry for quantum logic gates. %\cite{brandl2017quantum}
    \item \emph{Quantum-classical computer interface}: the hardware and software that provides the boundary between classical computers and the quantum computing layer.
    \item \emph{Quantum programming environment}: quantum programming languages and development environment. % provides first an IDE as well as a simulate support, second a quantum assembly language to instruct the QPU and finally a programming abstraction to write a hight-level quantum programs \cite{DBLP:journals/corr/abs-1803-07407}.
    \item \emph{Business applications}: quantum software applications (based on quantum programming languages) that meet specific business requirements %\Foutse{its unclear why this is quantum...since it runs on the classical computer layer!}.
\end{itemize}

\subsection{Quantum Programming} %\heng{to revisit}

Quantum computing as a new general paradigm can massively influence how software is developed~\cite{piattini2020talavera,zhao2020quantum,garhwal2019quantum}. %This has gave birth to new specifics and software engineering approaches~\cite{piattini2020talavera,zhao2020quantum}.
Quantum programming is the process to design an executable quantum program to accomplish a specific task~\cite{garhwal2019quantum}. 
%Every block of code is composed of classical and quantum operations \cite{zhao2020quantum}. Classical operations act on classical bits in order to register the states and measurements of Qubits~%\heng{do we use qubits or Qubits?}; 
%quantum operations operate on the quantum computers using registers of Qubits. 
Quantum programming uses syntax-based notations to represent and operate quantum circuits and gates. 
Early efforts of quantum programming language development focused on the quantum Turing machine~\cite{doi:10.1098/rspa.1985.0070} but did not produce practical quantum programming languages.
Later efforts have turned to the quantum circuits model where the quantum system is controlled by a classical computer~\cite{osti_366453}.
This concept has given birth to many new quantum programming languages such as qGCL~\cite{SandersJW:quap}, LanQ~\cite{doi:10.1142/S0219749908004031}, Q\#~\cite{10.1145/3183895.3183901} and Qiskit \cite{Qiskit}. Prior work conducted extensive exploration along the lines of quantum programming~\cite{garhwal2019quantum} and quantum software development environments~\cite{larose2019overview}. The survey~\cite{zhao2020quantum} also provides a comprehensive overview of research works along these lines.

%like classical computing , QSE is not limited to quantum programming, quantum software development methods are thriving.To overcome the challenges in the software development and insure the high quality a series of steps are followed in the shape of a life cycle know as  Quantum Software Life Cycle (QSDLC)[23]:Software requirement analysis, software design ,software implementation, software testing and software maintenance.The model begins with requirement analysis step were developers discuss the requirement to be developed to achieve their goal.In fact the scoop of the work is define along with the requirement that are going to be satisfied.The models follow with the design step, this is where the architectural and detailed design is made [75].At the implementation step the developer start coding following the requirement and design agreed on in the previous steps.To defect in the software and verify the behavior of the software the testing step come to action before releasing the system.Finally as the last step, maintenance represents the changes and the modification after the release of the quantum software.[3 23 24]

\vspace{-1mm}
\subsection{Quantum Software Engineering}
Quantum software engineering (QSE) is still in its infancy. 
As the result of the first International Workshop on Quantum Software Engineering \& Programming (QANSWER), researchers and practitioners proposed the ``Talavera Manifesto'' for quantum software engineering and programming, which defines a set of principles about QSE~\cite{piattini2020talavera}, including: % (e.g.,``QSE is agnostic regarding quantum programming languages and technologie'')~\cite{piattini2020talavera}.  
\emph{(1) QSE is agnostic regarding quantum programming languages and technologies; (2) QSE embraces the coexistence of classical and quantum computing; (3) QSE supports the management of quantum software development projects; (4) QSE considers the evolution of quantum software; (5) QSE aims at delivering quantum programs with desirable zero defects; (6) QSE assures the quality of quantum software; (7) QSE promotes quantum software reuse; (8) QSE addresses security and privacy by design; and (9) QSE covers the governance and management of software}.

Zhao~\cite{zhao2020quantum} performed a comprehensive survey of the existing technology in various phases of quantum software life cycle, including requirement analysis, design, implementation, testing, and maintenance. 
Prior work~\cite{moguel2020roadmap, piattini2020quantum, barbosa2020software, piattini2021toward} also discussed challenges and potential directions in QSE research, such as modeling~\cite{barbosa2020software} and quantum software processes \& methodologies~\cite{moguel2020roadmap}, and design of quantum hybrid systems~\cite{piattini2021toward}. 
%In addition, prior work conducted extensive exploration along the lines of quantum software programming~\cite{garhwal2019quantum} and quantum software development environments~\cite{larose2019overview}. The survey~\cite{zhao2020quantum} provides an comprehensive overview of the work along these lines.
Different from prior work, this work makes the first attempt to understand the challenges of QSE perceived by practitioners.

\vspace{-1mm}
\subsection{Topic Analysis of Technical Q\&As}
Prior work performs rich studies on technical Q\&A data, especially on Stack Exchange data\cite{StackExchange.data}. Here we focus on prior work that performs topic analysis on technical Q\&A data. 
Topic models are used extensively in prior work to understand the topics of general Stack Overflow posts and the topic trends~\cite{wang2013empirical,chen2019modeling, barbosa2020software,allamanis2013and}. 
Prior work also leverages topic models to understand the topics of Stack Overflow posts related to specific application development domains, such as mobile application development~\cite{linares2013exploratory, rosen2016mobile}, client application development~\cite{venkatesh2016client}, machine learning application development~\cite{alshangiti2019developing}, as well as concurrency~\cite{ahmed2018concurrency} and security~\cite{yang2016security} related development.
In addition, prior work leverages topic models to understand non-functional requirements communicated in Stack Overflow posts~\cite{zou2017towards,zou2015non}.
Zhang et al.~\cite{zhang2015multi} use topic models to detect duplicate questions in Stack Overflow. 
Finally, Treude et al.~\cite{treude2019predicting} proposes an automated approach to suggest configurations of topic models for Stack Overflow data.
Most of these studies use the Latent Dirichlet Allocation (LDA) algorithm or its variants to extract topics from the technical Q\&A data. 
In this work, we also leverage the widely used LDA algorithm to extract topics from the technical Q\&A data related to quantum software enginering.

% Understanding topics and trends
% An Empirical Study on Developer Interactions in StackOverflow
% Modeling stack overflow tags and topics as a hierarchy of concepts
% What are developers talking about? An analysis of topics and trends in Stack Overflow
% Why, When, and What: Analyzing Stack Overflow Questions by Topic, Type, and Code

% Topics of Mobile application development related questions
% An Exploratory Analysis of Mobile Development Issues using Stack Overflow
% What are mobile developers asking about? A large scale study using stack overflow

% Topics of Client application development related questions
% What Do Client Developers Concern When Using Web APIs? An Empirical Study on Developer Forums and Stack Overflow

% Concurrency development
% What Do Concurrency Developers Ask About? A Large-scale Study Using Stack Overflow

% Security related questions
% What Security Questions Do Developers Ask? A Large-Scale Study of Stack Overflow Posts

% Machine learning application development
% Why is Developing Machine Learning Applications Challenging? A Study on Stack Overflow Posts

% Duplicate Question Detection
% Multi-Factor Duplicate Question Detection in Stack Overflow

% Configurations for Topic models on Stack Overflow
% Predicting Good Configurations for GitHub and Stack Overflow Topic Models

% Understanding NFRs
% Towards comprehending the non-functional requirements through Developers’ eyes: An exploration of Stack Overflow using topic analysis
% Which Non-functional Requirements do Developers Focus on?
\vspace{-1mm}
\subsection{Topic Analysis of Issue Reports}
Issue reports have been widely explored in prior work. Here we focus on studies that apply topic analysis on issue report data. 
Prior work leverages topic models to automatically assign issue reports to developers (\emph{a.k.a.} bug triage)~\cite{zhang2014novel,xia2013accurate,naguib2013bug,xia2016improving}. These studies first uses topic models to categorize the textual information in the issue reports, then learn mappings between the categorized textual information and developers.
Prior work also leverages topic models to automatically detect duplicate issue reports based on the similarity of their topics~\cite{hindle2016contextual,nguyen2012duplicate,zou2016duplication}.
Nguyen et al.~\cite{nguyen2011topic} use topic models to associate issue reports and source code based on their similarities, in order to help developers narrow down the searched source code space when resolving an issue. 
Finally, prior work also studies the trends of topics in issue reports~\cite{martie2012trendy, aggarwal2014mining}.
LDA and its variants are the most popular topic modeling approaches used in these studies.
Therefore, we also leverage LDA to extract topics from GitHub issue reports related to QSE. %quantum software engineering.

% Automated bug triage
% A Novel Developer Ranking Algorithm for Automatic Bug Triage Using Topic Model and Developer Relations
% Accurate Developer Recommendation for Bug Resolution
% Bug report assignee recommendation using activity profiles
% DRETOM- developer recommendation based on topic models for bug resolution
% Improving Automated Bug Triaging with Specialized Topic Model

% duplicate issue report detection
% A contextual approach towards more accurate duplicate bug report detection and ranking
% Duplicate bug report detection with a combination of information retrieval and topic modeling
% Duplication Detection for Software Bug Reports based on Topic Model

% Bug localization
% A Topic-based Approach for Narrowing the Search Space of Buggy Files from a Bug Report

% Evaluating issue report quality
% Mining the coherence of GNOME bug reports with statistical topic models

% Issue prioritization
% Towards Prioritizing GitHub Issues

% Understanding topic trend
% Trendy bugs- Topic trends in the Android bug reports
% Mining issue tracking systems using topic models for trend analysis, corpus exploration, and understanding evolution

%% file: Sections/experimentSetup.tex
\section{Experiment Setup} \label{sec:setup}

%\heng{Need a space after each punctuation (, and . and others)}

%The section illustrate the methodology we followed in this work. In order to answer the research questions from RQ1 to RQ3. Starting from the data collection, data processing until data quantitative and qualitative analysis. The methodology is resumed in the figure 2 bellow. 
This section describes the design of our empirical study. %data collection and data processing.
\vspace{-1mm}
\subsection{Overview}
\begin{figure}[!t]
\vspace{-5mm}
    \centering
    \includegraphics[width=0.49\textwidth]{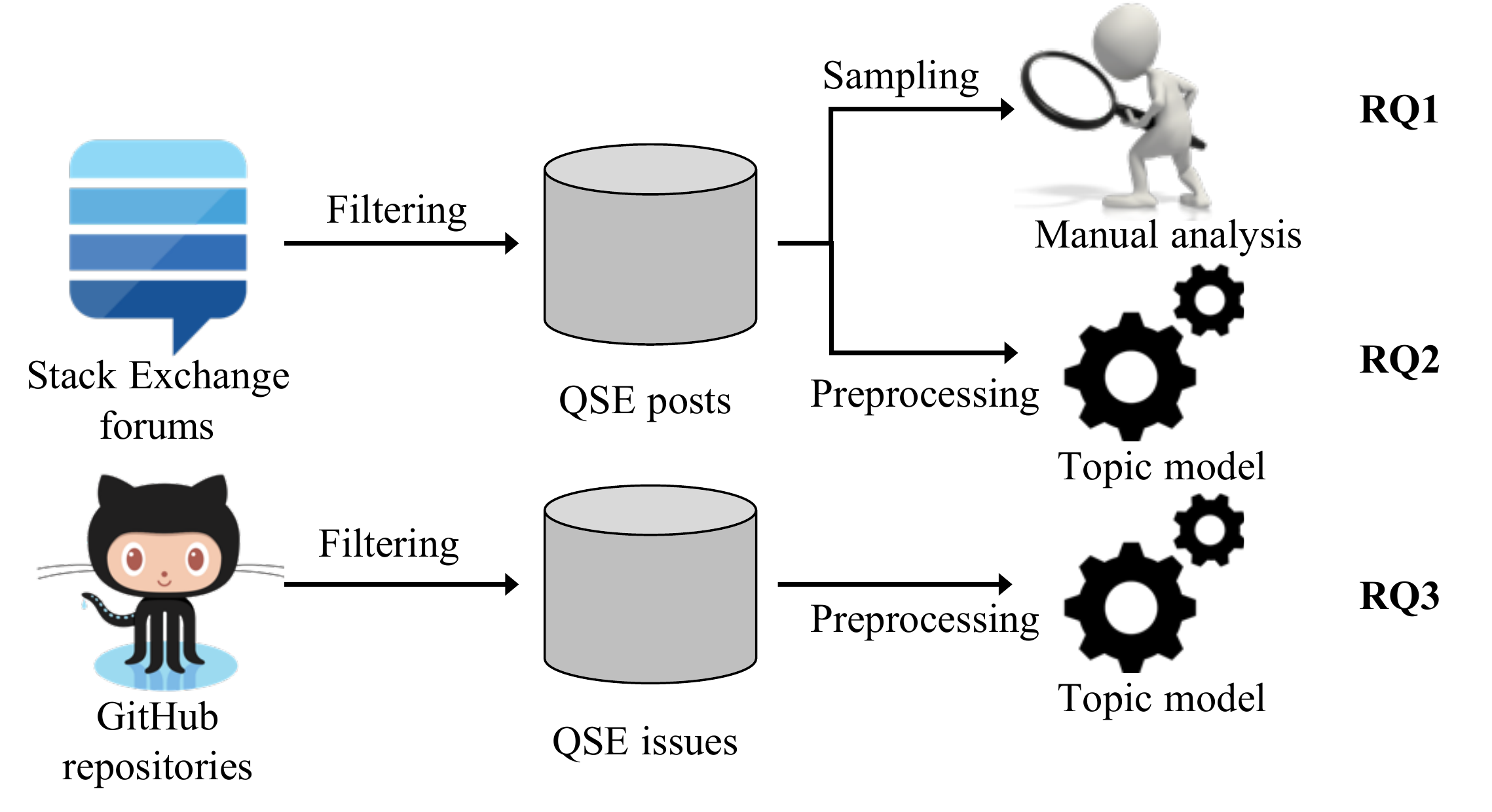}
    \vspace{-5mm}
    \caption{Overview of our empirical study}
    \label{fig:overview}
    \vspace{-5mm}
\end{figure}

Figure~\ref{fig:overview} provides an overview of our empirical study. 
We study QSE-related posts on Stack Exchange (SE) forums and the issue reports of quantum computing GitHub projects.
From Stack Exchange forums, we first use tags to filter QSE-related posts. In RQ1, we manually analyze a statistically representative sample of these posts to understand the type of information sought by developers. %and the categories \Foutse{of issues contained in these posts?} of these posts. 
In RQ2, we use automated topic models to analyze the topics of these posts and their characteristics.
From GitHub repositories, we first apply a set of heuristic rules to filter the quantum computing projects. Then we extract the issue reports of these quantum computing projects. Finally, we perform topics modeling on these issue reports to analyze the topics in the textual information of the issue reports (RQ3). 
We describe the details of our data collection and analysis approaches in the rest of this section.

\vspace{-1mm}
\subsection{Stack Exchange forums data collection }\label{AA}
%\heng{To be on the same page that Stack Overflow is also a Stack Exchange forum.}
We follow three steps to collect QSE related data from Stack Exchange forums. 
First, %we select the Stack Exchange forums that include QSE related posts. 
we collect Q\&A data from four Stack Exchange forums.
Second, we identify a set of tags that are related to QSE. Finally, we use the identified tags to select the posts that are related to QSE. We explain the steps below.

%\heng{to check below}
\noindent \textbf{Step 1: Collecting technical Q\&A data.} 
We extract technical Q\&A data from four Stack Exchange forums: Stack Overflow\cite{stackoverflow}, Quantum Computing Stack Exchange\cite{quantumcomputing.stackexchange}, Computer Science Stack Exchange\cite{cs.stackexchange}, and Artificial Intelligence Stack Exchange\cite{ai.stackexchange}.   
We consider the Stack Overflow forum as it contains posts related to quantum programming and it is widely used for studying various software engineering topics (e.g., mobile app development~\cite{linares2013exploratory, rosen2016mobile}, machine learning application development~\cite{alshangiti2019developing}, etc.).
We consider the other three forums because they contain posts that discuss topics related to quantum computing and quantum programming.
%for our work we chose stackoverflow(SO), QuantumComputing.stackExchange, ai.stackexchange and cs.stackexchange. 
%Since we are focusing on studying the challenges of Quantum computing and Software engineering, looking into SO is a common sense because SO is a popular Q\&A for developer to discuss topics related to software engineering~\cite{uddin2021understanding}. 
We extracted the post data from these forums with the help of the Stack Exchange Data Explorer\cite{data.stackexchange}.
%to end up with the most up to date data for our study. 
Stack Exchange data explorer holds an up to date data for these forum between 08-2008 and 03-2021.
%In total, we extracted 52,675,347 \heng{is the number for SO only or for all four forums?} posts from these forums between \heng{month} 2008 and \heng{month} 2021. The detailed number of extracted posts for each forum is shown in Table~\ref{tab:tags}\heng{too add}.
%In addition to SO, Quantum Computing artificial intelligence and computer science forums of StackExchange were explored in order to make sure we are mining most of the data.

%The final extracted data $D_{init}$ contains the following information: first the posts title and body coming with text and code examples. Second, a flag indicate if the post has an accepted answer. Third, creation and edition date. More over we had the user ID that represent the post owner and finally user-provided tags associated to the posts. Each question can have 1-5 tags.

%\textbf{Step 2: Quantum computing Tags selection: }
\noindent \textbf{Step 2: Identifying tags related to QSE.}
%Not all the posts StackExhchange are related to quantum computing. Therefore we have to extract the relevant posts for our study. 
%We first select the user-defined tags relevant to QSE. Then we use these selected tags to filter the posts that are related to QSE. 
The studied Stack Exchange forums use user defined tags to categorize questions. 
We follow two sub-steps to select the tags that are related to QSE.
%\begin{itemize}
%\item %\heng{This paragraph needs to be more specific: What is the initial searched keywords? are they searched on all four forums? How many most relevant posts are inspected in the initial round? how many tags in the initial set? Please make the process clear and precise} 
%We initially looked for posts related to quantum computing in SO because we are only interested in quantum software engineering tags. Therefore
We started by searching for questions with the tag ``quantum-computing'' in the entire Stack Exchange dataset $D$ through the data exchange explorer. We obtained 254 questions tagged with ``quantum-computing'' from the studied forums. After manually inspecting the 30 most voted questions, we selected an initial tag set $T_{init}$ consisting of ten tags including ``quantum-computing'', ``qiskit'', ``qsharp'', ``q\#'', ``quantum-development'', ``quantum-circuit'', ``ibmq'', ``quantum-ai'', ``qubit'' and ``qutip''. 
Then we extracted the questions related to $T_{init}$ from the initial dataset $D$ and obtained a new set of questions $P$. In order to expand the initial tag set, we extracted the frequently co-occurring tags with $T_{init}$ from $P$ and build a new tag set $T_{2}$. 

%\Foutse{how exactly? usually we look at tags frequently co-occurring with those from the initial set of tag, to build this new tag set...is that what we did? We should clarify how we created this new tag set.}
%$D$ \heng{not clear what is the initial dataset} 

%\item 
Not all the tags in $T_{2}$ are related to quantum computing. %There may be tags related to software engineering (i.e, algorithm, error-correction, CPU, hard-drive). 
To determine the final tag set $T_{final}$, following previous work~\cite{uddin2021understanding} \cite{9240667}, we filter the tags in $T_{2}$ based on their relationships when the initial tag set $T_{init}$. For each tag t in $T_{2}$, we calculate: %\heng{I feel the equations for significance and relevance should be switched, double check}
\begin{equation}
(\textnormal{Significance})\,\alpha (t) = %\frac{No\,\,of\,\,questions\,\,with\,\,tag\,\,t\,\,in\,\,P}{No\,\,of\,\,questions\,\,with\,\,tag\,\,t\,\,in\,\,D}
\frac{\textnormal{\# of questions with tag } t \textnormal{ in } P}{\textnormal{\# of questions with tag } t \textnormal{ in } D}
\end{equation}
\begin{equation}
(\textnormal{Relevance})\,\beta (t) = \frac{\textnormal{\# of questions with tag } t \textnormal{ in } P}{\textnormal{\# of questions in } P}
%\frac{No\,\,of\,\,questions\,\,with\,\,tag\,\,t\,\,in\,\,P}{No\,\,of\,\,questions\,\,P}
\end{equation}
To select a tag t, the value of significance-relevance $\alpha(t)$, $\beta(t)$ need to be higher than a threshold we set. To select the optimal threshold values for $\alpha$ and $\beta$, we experimented with a set of values respectively between 0.05, 0.35 and 0.001, 0.03. For each $\alpha$ and $\beta$ and for each tag above the threshold, we inspected the top 10 most voted posts and verified if the tag is related to QSE, we ended up with the optimal threshold
%\heng{how did you determine it's optimal}  
respectively equal to 0.005 and 0.2 which are consistent with previous work~\cite{10.1109/MSR.2017.5}~\cite{9240667}. The final tag set $T_{final}$ is formed of 37 tags in total. Since quantum computing is a wide topic and our focus is QSE, we further manually inspected the description of each tag t in $T_{final}$ and the top 10 questions of each tag in each studied forum to remove tags that are not related to QSE. Finally our tag set $T_{final}$ was reduced from 37 to 18 tags (14 unique tags as different forums have tags with the same names).
Table~\ref{tab:tags} lists our final set of tags.
%\end{itemize}
\input{tables/tagSetTable}

\noindent\textbf{Step 3: Selecting questions and answers.}
We extract the final sets of questions and answers using the final tag sets shown in Table~\ref{tab:tags}. We select all the posts that are tagged with at least one of the tags. %Only the tagged questions with one of our tag t in $T_{final}$ are considered. 
We ended up with a total of 3,117 questions and answers 
%\heng{is the number for the number of questions/posts, or question and answers?} 
from the four considered forums in our data set $D_{final}$.
%coming from SO and SE.
35\% of the final data are answers where 65\% are questions. 
The number of posts (questions and answers) extracted from each forum is shown in Table~\ref{tab:tags}.
%\heng{too add}.
%\heng{we need to make it clear what is a post: is it a q\&a thread, or just one question or one answer?}
\vspace{-1mm}
\subsection{GitHub issues data collection}
%The GitHub data is easy to obtain and analyze with the help of of GitHub Rest API~\cite{githubrestapi}. We can go through all the necessary information like the commit messages, pull request messages and issues, etc. For the sake of our study, we crawled the issues related to a set of quantum computing project. The GitHub data set was downloaded in March 2021. To collect the GitHub issues we followed two steps project selection.
In this work, we study the issue reports of quantum computing projects on GitHub. We downloaded the GitHub selected quantum computing projects issues in March 2021.
%\heng{be more specific: what GitHub data set, may be give a link?}.
We follow three steps described below to extract the issue reports of quantum computing projects from GitHub.

\noindent\textbf{Step 1: Searching candidate projects.}
%The first step is to select Quantum computing projects. Different famous Quantum framework such as "Qiskit" and "q\#" are shared in GitHub as it host a large number of open source projects. 
We search for quantum computing related projects using three criteria:
%\begin{itemize}
1) The description of the project must be in English (i.e., for us to better understand the content).
2) The project name or description must contain the word ``quantum'' (the word quantum is case sensitive in the project name or description).
3) The project is in a mainline repository (i.e., not a fork of another repository).
%\end{itemize}
We end up with a total of 1,364 repositories.

\noindent\textbf{Step 2: Filtering quantum computing projects.}
We filter the searching result and identify quantum computing related projects with three criteria:
%\begin{itemize}
1) To avoid selecting student assignments, following previous work~\cite{Businge2018CloneBasedVM}~\cite{Businge2019StudyingAA}, we select repositories that were forked at least two times. %We remove 803 projects.
2) The projects must have a sufficient history of development for us to analyze the issue reports. Therefore, we select the projects that were created at least 10 months earlier than the data extraction date. Moreover, only the projects that have at least 100 commits and 10 issues are selected.
%\Foutse{so are we consider projects with one release or not? we say newly released projects can't have issues...that is not correct...what is a newly released project here, exactly?} can not have issues and that is what we are looking to study. %\heng{cannot understand this sentence}. %These criteria only removed another 417 projects.
3) To ensure the quality of the project selected, we manually inspect the projects' descriptions and remove projects that are not related to quantum computing, projects that are created %or those 
for hosting quantum computing related documentation, as well as lecture notes related to quantum computing. % and discarded 13 further repositories about quantum documentation and lecture note. More over, we find 9 another project not related to quantum computing like 'foxyproxy/firefox-extension'.
%\end{itemize}
Finally, we obtain a total of 122 projects directly related to quantum computing applications. 

\noindent\textbf{Step 3: Extracting issue reports.}
We use the GitHub Rest API~\cite{githubrestapi} to extract all the issue reports of the final 122 projects on GitHub. In total, we obtain 43,979 issue reports. %from these projects.
\vspace{-1mm}
\subsection{Data pre-processing for topic modeling}

%\heng{you mentioned “For the Q\&A posts, we join the title and the text body of the questions to create one final body The text body.” Did you consider answer in topic modeling and how did you do for answers? For github issue ports, what parts of text did you use? did you join title and description? did you consider comments? Please make the following steps clear}
%\heng{divide the process into two paragraphs, one for Q\&A and another for GitHub issues.}
%For the topic modeling analysis, to filter out the noise in the text body of the Q\&A posts and GitHub issues ,We processed the data as follow:
We build one topic model on the Stack Exchange forum data and another topic model on the GitHub issue data. Below we describe how we pre-process these two types of data before feeding them into topic models.

%\heng{Is it right? $\rightarrow$ mohamed says : yes }
\noindent\textbf{Pre-processing Q\&A post data.}
We treat each post (i.e., a question or an answer) as an individual document in the topic model. For each question, we join the title and the body of the question to create a single document.
As Q\&A posts contain code snippets between \textless code\textgreater { and} \textless /code\textgreater { which} may bring noise to our topic models, we remove all text between \textless code\textgreater { and} \textless /code\textgreater.
We also remove HTML tags (e.g., \textless p\textgreater\textless /p\textgreater), URLs and images from each post.
In addition, we remove stop words (e.g., ``like'', ``this'', ``the''), punctuation, and non-alphabetical characters using the Mallet and NLTK stop words set.
Finally, we apply the Porter stemming \cite{potterStemming} to normalize the words into their base forms (e.g., ``computing'' is transformed to ``comput''), which can reduce the dimensionality of the word space and improve the performance of topic models~\cite{PerformanceAnalysis:Stemming}%\heng{any citation?}. 
%The morphological form improves the contextual understanding by strengthening the similarity while conserving the diversity. Besides, the stemming will reduce the dimensionality, as a result, enhance the performance of the Topic modeling.

\begin{comment}
\noindent \textbf{Technical forums posts:}
\begin{enumerate}
\item We join the title and the text body of the questions to create one final body
The text body. Also Q\&A posts contains code snippets between \textless code\textgreater\textless \textbackslash code\textgreater which contains programming languages code that can be irrelevant to our analysis and can lead to bad topic modeling. We remove every text between \textless code\textgreater.
\item HTML tags like \textless p\textgreater\textless \textbackslash p\textgreater, URLs and images are removed from Q\&A posts final body.
\item We remove the stop word such as like, this, the, punctuation and non-alphabetical characters using Mallet  \cite{McCallumMALLET} and NLTK stop words set.
\item We apply Porter stemming \cite{potterStemming} to normalize the word and return it to the base form (i,e 'computing', transformed to 'comput'). The morphological form improves the contextual understanding by strengthening the similarity while conserving the diversity. Besides, the stemming will reduce the dimensionality, as a result, enhance the performance of the Topic modeling.
\end{enumerate}
\end{comment}

%\noindent \textbf{GitHub issues:}
\noindent\textbf{Pre-processing issue report data.} 
We treat each issue report as an individual document in the topic model. %For each issue report, 
We join the title and the body of each issue as a single document.
Similarly, we remove code snippets, URLs and images from the issue body.
Since there are no tags in GitHub issues that identify code snippets, we look for backquote ” ” or triple backticks ``` in the content of the issues and remove the code enclosed between this punctuation. 
We also remove stop words, non-alphabetical characters, and punctuation.
Finally, we apply Porter stemming to normalize the words into their base forms.

\begin{comment}
\begin{enumerate}
    \item We join the title and the body of the issue
    \item Since there is no HTML tags in GitHub issues, to remove the code snippets we look for back quote ” ” in the content of the issues and remove the code enclosed between this punctuation. 
    \item URLs, numbers, punctuation, non-alphabetical characters and stop words are removed from final body.
    \item We apply Porter stemming to normalize the word and transform it to the base form.
\end{enumerate}
\end{comment}

\vspace{-1mm}
\subsection{Topic modeling}
%Our study involves finding discussed topics related to Quantum software engineering from both Q\&A forums and GitHub platform. Because the tags in GitHub are not expressive and fail to deliver a descriptive information about what is discussed in Q\&A forums and GitHub issues does not have Tags, 
We use automated topic modeling to analyze the topics in the Q\&A posts and issue reports.
Specifically, we use the Latent Dirichlet Allocation (LDA) algorithm~\cite{blei2003latent} to extract the topics from both of our datasets. LDA is a probabilistic topic modeling technique that derives the probability distribution of frequently co-occurred word sets (i.e., topics) in a text corpus.
A topic is represented by a probability distribution of a set of words, while a document is represented as a probability distribution of a set of topics.
LDA is widely used for modeling topics in software repositories~\cite{chen2016survey}, including technical Q\&A posts (e.g,~\cite{barua2014developers}) and issue reports (e.g.,~\cite{hindle2016contextual}).
%The posts and the issues are separately grouped into k number of topics after I iteration. 
We use two separate topic models to extract the topics from the Q\&A post data and the issue report data.
For a better performance of the topic modeling and a good classification quality, following previous work~\cite{9240667}~\cite{inproceedings}, we consider both uni-gram and bi-gram of words in our topic models.

\noindent\textbf{LDA Implementation.} 
We use the Python implementation of the Mallet topic modeling package~\cite{McCallumMALLET} to perform our topic modeling.
The Mallet package implements the Gibbs sampling LDA algorithm and uses efficient hyper-parameter optimization to improve the quality of the derived topics~\cite{McCallumMALLET}.

%\noindent\textbf{Determining the number of topics $K$.}
\noindent\textbf{Determining topic modeling parameters.}
%\heng{topic modeling parameters}
The number of topics ($K$) is usually manually set by the user as it controls the granularity of the topics~\cite{9240667}. 
The $\alpha$ parameter controls the topic distribution in the documents (i.e., Q\&A posts or issue reports), while the $\beta$ parameter controls the word distribution in the topics.
In this work, we use the topic coherence score~\cite{Rder2015ExploringTS} to evaluate the quality of the resulting topics and determine the appropriate parameters ($K$, $\alpha$, and $\beta$), similar to prior work~\cite{9240667, 10.1145/3183895.3183901}. 
The coherence score measures the quality of a topic by measuring the semantic similarity between the top words in the topic. Thus, this score distinguishes between topics that are semantically interpretable and topics that are coincidences of statistical inference \cite{Rder2015ExploringTS}.
Specifically, we use the Gensim Python package's \texttt{CoherenceModel} \cite{Gensim.CoherenceModel} module to calculate the coherence scores of the resulting topics. 
To capture a wide range of parameters and keep the topics distinct from each other, we experiment with different combination of the parameters, by varying the values of $K$ from 5 to 30 incremented by 1 each time, the values of document-topic distribution $\alpha$ from 0.01 to 1 incremented by 0.01~\cite{Han2020WhatDP}, and the values of word-topic distribution $\beta$ from 0.01 to 1 incremented by 0.01~\cite{Han2020WhatDP}. We retain the resulting topics with the highest average coherence score. 
%\heng{average?}  

After getting the automatically derived topics, we manually analyze the resulting topics and assign meaningful labels to the topics. We elaborate more on this process %, which is described 
in RQ2 and RQ3 for the Q\&A post topics and the issue report topics, respectively.
\begin{comment}
%\heng{move the following to RQ 2\&3 approaches if not done yet.}
\Foutse{the following paragraph should be moved to the approach of the RQs!}
To analyze the discussions in Q\&A and GitHub platforms and capture the topics. We run LDA algorithm on the different data set corpus. The data set is composed of two corpus: 43,979 GitHub issues and 3,117 Q\&A posts. Since LDA has no semantic knowledge, the topic has to be manually categorized to give a meaning.
\end{comment}

\begin{comment}
\subsection{Type of questions manual labeling}
To detect the type of question in QSE topics, following previous works~\cite{beyer2020kind} we perform manual analysis on statistical significant random sample SO and SE questions. On two round, and using cart sorting technique the authors label the posts according to 9 categories: API usage, Theoretical, Errors, Conceptual,  Discrepancy, Learning, Review, Tooling and API change.
\end{comment}

%% file: tables/tagSetTable.tex
\begin{table}[!t]
\vspace{-5mm}
\centering
\caption{Our selected tags and the number of questions and answers}
\vspace{-5pt}
\resizebox{\columnwidth}{!}{%
\begin{tabular}{@{}llcc@{}}
\rowcolor[HTML]{EFEFEF} 
\toprule
\textbf{Stack Ex. forum} &
  \textbf{Tag set} &
  \multicolumn{1}{l}{\textbf{\#Q}} &
  \multicolumn{1}{l}{\textbf{\#A}} \\ \midrule
Stack overflow &
  \begin{tabular}[c]{@{}l@{}}post-quantum-cryptography, q\#, \\ quantum-computing, qiskit, qcl, \\ qutip, qubit, tensorflow-quantum\end{tabular} &
  250 &
  183 \\ \midrule
\begin{tabular}[c]{@{}l@{}}Quantum computing\end{tabular} &
  \begin{tabular}[c]{@{}l@{}}programming, classicalcomputing, \\ q\#, qiskit, cirq, ibm-q-experience, \\ machine-learning, qutip\end{tabular} &
  1534 &
  778 \\ \midrule
\begin{tabular}[c]{@{}l@{}}Computer science\end{tabular} &
  quantum-computing &
  238 &
  117 \\ \midrule
\begin{tabular}[c]{@{}l@{}}Artificial intelligence\end{tabular} &
  quantum-computing &
  13 &
  4 \\ \bottomrule
\end{tabular}
}
\vspace{-5mm}
\label{tab:tags}
\end{table}

%% file: Sections/result.tex
\section{Experiment Results} \label{sec:results}
In this section we report and discuss the results of our three research questions. For each research question, we first present the motivation and approach, then discuss the results for answering the research question.

\input{Sections/RQ1}

\input{Sections/RQ2}

\input{Sections/RQ3}

%% file: Sections/RQ1.tex
%\subsection*{\textbf{RQ1: Categories of QSE questions on technical Q\&A forums}}
\vspace{-3pt}
\subsection*{\textbf{RQ1: What types of QSE questions are asked on technical forums?}}
\subsubsection{\textbf{Motivation}}
%Quantum computing is an emerging and under-explored area, particularly from the perspective of software engineering.
In order to understand QSE challenges developers are facing, we first want to understand what types of questions they are asking (e.g., whether they are asking questions about using APIs or fixing errors). This is important to identify the areas in which QSE developers should be supported and the type of resources that they need.
%Thus, as an initial attempt to understand quantum software engineering and its challenges, we want to investigate what categories of questions people are asking. 
%Thus, in this RQ, we performed a manual analysis on a sample of QSE related questions from our studied technical forums.
Similar to prior work~\cite{beyer2020kind}, we focus on the intent behind the questions asked by QSE developers instead of the topics of the questions. %we focus on developers' intent behind asking a question instead of the topics of the questions.

\subsubsection{\textbf{Approach}}

To identify the type of questions that users are asking in technical forums, we performed a manual analysis of a statistically representative sample from our studied QSE questions. We sampled 323 questions with a confidence level of 95\% and a confidence interval of 5\%. For each question, we examined its title and body, to understand the intent of the user who posted the question. 
We used a hybrid card sorting approach to perform the manual analysis and assign labels (i.e., types of questions) to each sampled question.
Specifically, we based our manual analysis on an existing taxonomy of the types of questions asked on Stack Overflow~\cite{beyer2020kind} and added new types when needed.
For each question we assigned one label; in case a question is associated with two or more labels, which we found only in a few cases, we chose the most relevant one. 

\noindent \textbf{Hybrid card sorting process.}
Two authors of the paper (i.e., coders) jointly performed the hybrid card sorting. 
We split the sampled data into two equal subsets and performed the sorting in two rounds, similar to prior work~\cite{li2020qualitative}. Our process guaranteed that each question is labelled by both coders.
\begin{enumerate}
    \item \textbf{First-round labeling.} Each coder labels a different half of the questions independently. %this step took 2-3 days.
    \item \textbf{First-round discussion.} In order to have a consistent labeling strategy, we had a meeting to discuss the labeling results in the first round and reached an agreed-upon set of labels. A third author of the paper is involved in the discussion.
    \item \textbf{Revising first-round labels.} Each coder updated the first round labeling results based on the discussion.
    \item \textbf{Second-round labeling.} Each coder labeled the other half of the questions independently based on the agreed-upon labels in the first round. New labels are allowed in this round. %Since the data was evenly sampled and distributed between the authors such that we are sure that the posts are labeled by two authors. This step took 2-3 days.
    \item \textbf{Second-round discussion.} We had a meeting to discuss the second-round labeling results, validate newly added labels and verify the consistency of our labels. A third author is also involved in the discussion.
    \item \textbf{Revising second-round labels.} Based on the second-round discussion, each coder revised the labels and finalized its individual labeling of the questions. We calculate the inter-coder agreement after this step.
    \item \textbf{Resolving disagreement.} We had a final meeting to resolve the disagreement in our labeling results and reached the final label for each question. For each difference in our labels, the two coders and a third author discussed the conflict and reached a consensus.
\end{enumerate}

\noindent \textbf{Inter-coder agreement. }
We measured the inter-coder agreement between the coders and obtained a Cohen’s kappa $k$ value of 0.73 which indicates a substantial agreement~\cite{MaryMcHugh}. Therefore our manual labeling results are reliable.

\subsubsection{\textbf{Results}}

\input{tables/QA_taxonomy}

Table \ref{table:intent} shows the result of our qualitative analysis for identifying the categories of questions in technical forums. Among the 323 questions we analyzed, we could not assign a label to only one question. In the table, we provide the description of each category and how frequent it appears in our qualitative analysis.

\noindent \textbf{All seven categories of Stack Overflow questions identified in prior work appear in QSE-related posts.}
Prior work~\cite{beyer2020kind} identified seven categories of questions on Stack Overflow by studying Android-related questions, including \texttt{API usage}, \texttt{Conceptual}, \texttt{Discrepancy}, \texttt{Errors}, \texttt{Review}, \texttt{API change}, and \texttt{Learning}, ordered by their occurrence frequency. 
Although quantum computing is still a new area, people start to ask all these different categories of questions, indicating that quantum computing face similar software engineering challenges (e.g., \texttt{API usage} and \texttt{API change}) as other software engineering domains. 
Similar to prior work, we find that \texttt{API usage} is the most frequent category with 26.3\% %\heng{change to 26.3\%, fix everywhere} 
instances.
The questions of this category are usually identified by ``how to''; e.g., %. For example, one question asks 
``\textit{How to return measurement probabilities from the QDK Full-state simulator?}''

\noindent \textbf{The categories of \texttt{Errors} and \texttt{Learning} are relatively more frequent in QSE-related questions than in the prior taxonomy of question categories~\cite{beyer2020kind}.}
Compare to prior work~\cite{beyer2020kind} on classifying Android-related questions, we find that \texttt{Errors} and \texttt{Learning} questions are relatively more frequent. 
As quantum computing is still an emerging domain, people practicing it face many errors when developing quantum computing applications and they find it challenging to find learning resource for quantum computing.
An example of the \texttt{Errors} category is ``\textit{I have Qiskit installed via Anaconda and a virtual environment set up in Python 3.8. ... I get an error. I'm not sure what the problem is. How do I fix it?}''.
Another example for the \texttt{Learning} category is ``\textit{How do I learn Q\#? What languages should I know prior to learning Q\#? How do I get started with quantum computing?}''.
These findings suggest the need to develop tools or resources to help developers avoid or address such errors, as well as developing tutorials, books, and other learning resources to help beginners get acquainted with quantum computing.

\noindent \textbf{Two new categories of questions (i.e., \texttt{Theoretical} and \texttt{Tooling}) emerge in QSE-related posts.}
In fact, the category of \texttt{Theoretical} is the second most frequent among all categories. This category is usually associated with keywords such as ``can someone explain'', ``what is'', and ``does quantum''. 
An example question of this category is ``What is the analysis of the Bell Inequality protocol in Cirq's `examples'?'' where Cirq\cite{cirq_developers_2021_4586899} is a Python library for developing quantum computing applications.
This category of questions indicates that people have challenges understanding the theoretical concepts behind quantum computing code. Future efforts are needed to explain such theoretical concepts for developers.
The category of \texttt{tooling} represents questions that are looking for tools, frameworks, or libraries that can help solve a QSE-related problem or verifying whether a tool, framework, or library can help solve a problem. For example, ``\textit{I want to use Blender and Blender Python Scripts working with Qiskit. How can I do this? How to make communication between Blender and Qiskit installed with Anaconda Python?}''.
This category indicates the lack of established tools for supporting quantum program development.

\begin{tcolorbox}
\vspace{-2mm}
We identified nine categories of QSE-related questions in Stack Exchange forums. The categories \texttt{Theoretical}, \texttt{Errors}, \texttt{Learning}, and \texttt{Tooling} are new or become more frequent in QSE-related questions.
Our results highlight the need for future efforts to support developers' quantum program development, in particular, to develop learning resources, to help developers fix errors, and to explain theory behind quantum computing code.
\vspace{-2mm}
\end{tcolorbox}

%% file: tables/QA_taxonomy.tex
\begin{table*}[tbp]
\vspace{-2mm}
\caption{A taxonomy of Question Categories which bases on and extends ~\cite{beyer2020kind}}
%\resizebox{\textwidth}{!}{%
\begin{tabular}{llc}
\hline
\rowcolor[HTML]{EFEFEF} 
\multicolumn{1}{l}{\textbf{Category}} &
  \textbf{Description} &
  \textbf{Freq} \\ \hline
API usage &
  \begin{tabular}[c]{@{}l@{}}Questions of this category are usually identified by ``how to'', i.e., how to use an API or how to implement a functionality.\end{tabular} &
  85 \\ \hline
Theoretical$^{\mathrm{*}}$ &
  \begin{tabular}[c]{@{}l@{}}This category of questioners ask about theoretical explanations of quantum programs, algorithms, and concepts.\end{tabular} &
  54 \\ \hline
Errors &
  \begin{tabular}[c]{@{}l@{}}This category of questions search for explanations and solutions of errors and exceptions when developing\\or executing quantum programs.\end{tabular} &
  49 \\ \hline
Conceptual &
  \begin{tabular}[c]{@{}l@{}}Questions in this category are related to the limitation, background and the underlying concept of an API. \end{tabular} &
  45 \\ \hline
Discrepancy &
  \begin{tabular}[c]{@{}l@{}}Question of this category usually ask for explanations or solutions for unexpected results \\(e.g., ``what is the problem'', ``why not work''.\end{tabular} &
  31 \\ \hline
Learning &
  \begin{tabular}[c]{@{}l@{}}Questions in this category are searching for learning resources such as documentation, research papers, tutorials, or websites.\end{tabular} &
  22 \\ \hline
Review &
  \begin{tabular}[c]{@{}l@{}}This category describes questions like: ``How/Why this is working?'' or ``Is there a better solution?''.\\ Generally, the questions in this category look for a better solution to a problem or for help reviewing the current solution. \end{tabular} &
  17 \\ \hline
Tooling$^{\mathrm{*}}$ &
  \begin{tabular}[c]{@{}l@{}}This category describes questions like ``I am looking for ...'', ``Is there a tool for ...''.\\ These questions search for tools to solve a specific problem or check the features of a tool. \end{tabular} &
  16 \\ \hline
API change &
  \begin{tabular}[c]{@{}l@{}}This category of questions concern about changes of an API and the associated compatibility issues and other implications. \end{tabular} &
  2 \\ \hline
\multicolumn{3}{l}{$^{\mathrm{*}}$Categories newly identified in QSE-related questions.}
\end{tabular}
%}
\label{table:intent}
\vspace{-3mm}
\end{table*}

%% file: Sections/RQ2.tex
%\subsection{\textbf{RQ2: QSE topics in technical Q\&A forums}}
\vspace{-5pt}
\subsection*{\textbf{RQ2: What QSE topics are raised in technical forums?}}
\setcounter{subsubsection}{0}
\subsubsection{\textbf{Motivation}}
Developers post QSE-related questions and answers on technical forums. Their posts may reflect their faced challenges when learning or developing quantum programs. To understand their faced challenges, we use topic models to extract the semantic topics in their posts and analyze the characteristics of these topics.

\subsubsection{\textbf{Approach}} %\leavevmode

%\heng{Use the following three sections to organize the approach section:}
\noindent \textbf{Topic assignment and frequency.} 
%For the topics to have a meaning, we need to manually give each one a significant name that summarize his underlying concept. 
The automated topic modeling generated nine topics and distribution of co-occurring words in each topic. 
We then manually assigned a meaningful label to each topic.
Following prior work\cite{Yang:Xin-Li:Lo, 10.1145/3338906.3338939, 9240667}, to assign a meaningful label to a topic, the first author first proposed labels using two pieces of information: (1) the topic's top 20 keywords, and (2) the top 10-15 most relevant questions associated with the topic. Then, three authors of the paper reviewed the labels in meetings and reassigned the labels when needed. %During the meetings,we discussed with each other, reviewed and assigned the appropriate label to each topic. 
%We also checked the possibility of merge some topics because some topics can be similar with different vocabulary which LDA considers different. For instance, to reach a final agreement we went through, at least, 10 iteration and at the end we obtained 9 different topics. 
%\heng{overall it's good. You mix the labelling of topics and counting the percentages. They are two things.}
We obtained a meaningful label for each of the nine topics at the end.
For each topic, we measure the percentage of the posts (i.e., frequency) that have it as the dominant topic (i.e., with the highest probability).

\noindent \textbf{Topic popularity.} To understand developers' attention towards each topic, following previous work \cite{Yang:Xin-Li:Lo, 10.1145/3338906.3338939, 9240667}, we measured three metrics for each topic: (1) the median number of views of the associated posts, (2) the median number of associated posts marked as \texttt{favorite}, and (3) the median score of the associated posts. For each topic, the associated posts refer to the posts that have it as the dominant topic.

\begin{comment}
\begin{itemize}
    \item We group the posts by the dominant topic label.
    \item We aggregate the number of views, number of posts and post score with median aggregation function.
\end{itemize}
Hence \textbf{the topic is popular if he has the most number of view and the highest number of favorite and score}.
\end{comment}
%\heng{Add more details about how you calculated the three: for each topic, you get all the posts with that topic, then calculate the median values...}. 
%\heng{Suggestion: don't use very long senteces. try to break it to small steps and be precise: First, for each topic, we extract the posts that are associated with the topic. Then, we measure the median view count, median scores, and the median number of favorites of these posts.}

\noindent \textbf{Topic difficulty.} 
%\heng{check comments for topic popularity: use short sentences and be precise}
In order to better understand the most challenging aspects for developers, we measure the difficulty of each topic in terms of how difficult it is for the associated posts to get accepted answers.
%In order to understand how difficult each topic to answer for quantum software engineers, 
Following prior work \cite{Yang:Xin-Li:Lo, 10.1145/3338906.3338939, 9240667 }, for each topic, we measure two metrics: (1) the percentage of the associated questions with no accepted answer, and (2) the median time required by the associated questions to get an accepted answer (only considering the ones with an accepted answer). For each topic, the associated questions refer to the questions that have the topic as the dominant topic.
\begin{comment}
\begin{itemize}
    \item We group the posts by dominant topic label
    \item (1) We compute the percentage of the number of topics with and without accepted answer.
    \item (1) We select only the percentage of posts with no accepted answer.
    \item (2) We compute the difference between the date of an accepted answers and the creation data in hour per post. 
    \item (2) We aggregate the median time required by the question to get an accepted answer with median aggregation function.
\end{itemize}
Hence \textbf{The topic is difficult if after a long waiting time he receives a few answers.}. 
\end{comment}
%\heng{If possible, always try to motivate your approach: In order to understand whether a topic a general or project specific, we count the number of projects that have each topic.}

%\noindent \textbf{Topic evolution.} How we calculate evolution. May be optional.

\subsubsection{\textbf{Results}} %\leavevmode

\input{tables/tableQAtopics}
\noindent \textbf{We derived nine topics that are discussed in QSE-related posts, including traditional software engineering topics (e.g., \texttt{environment management} and \texttt{dependency management}) and QSE-specific topics (e.g., \texttt{quantum execution results} and \texttt{Quantum circuits})}.
Table \ref{table:QA_topic} describes the nine topics and their frequency in the analyzed posts. %As one can observe in QSE there is no wide range of topics discussed with a total of 9.Since the number of the detected topics in not big, We present a low level of granularity.Also for each topics we illustrate the percentage of the question asked order by their occurrence.The percentage indicate the dominance of a topic compared to others. 
Table~\ref{table:QA popularity} shows the median views, scores, and favorites of the posts associated with these topics.
The three most dominant topics are \texttt{environment management}, \texttt{dependency management}, and \texttt{algorithm complexity}.

\textbf{\texttt{Environment management}} is the most dominant topic representing 15.03\% of posts.
%\heng{or posts?}. 
For example, the most viewed question of this topic is %$Q_{18041396}$ 
``\textit{I downloaded the Quipper package but I have not been able to get haskell to recognize where all of the modules and files are and how to properly link everything}'' which gained 2772 views. %After 6 years this question did not receie any answers. 
%More over, from inspecting the three most viewed question for this topics 
%$Q_{19761700}$
Other examples include ``How can I run QCL (quantum programming language) on Windows?'' and %$Q_{2195}$ 
``Visualization of Quantum Circuits when using IBM QISKit''. 
%$Q_{53929599}$ 
%``Cannot import Aer from Qiskit 0.7''.
We can observe that users are new to quantum computing and facing problem while setting up their environment and installing their tools. 
This topic is also linked to the question category \texttt{tooling} that we derived in RQ1. 

\textbf{\texttt{Dependency management}} is the second most discussed topic representing 14.82\% of the posts. 
For example, the most viewed question (with 2,239 views) of this topic is %$Q_{54503699}$ 
``\textit{When trying the above code, I am receiving the following error: ModuleNotFoundError: No module named qiskit}'' where \texttt{qiskit} is an open source framework for quantum program development\cite{Qiskit}. We noticed that a large number of questions are directly related to \texttt{qiskit}. This can be explained by the lack of documentation or tutorials in using this framework. %The topic is also linked to the questions category \texttt{Discrepancy} and \texttt{Errors} derived from RQ1.

\textbf{\texttt{Algorithm complexity}} is the third most dominant topic. % in QSE with 14\% of the posts. 
This topic is about understanding the complexity of quantum algorithms and how to optimize quantum algorithms. For example, the most viewed question of this topic is %$Q_{16684}$ 
``\textit{For the other algorithms, I was able to find specific equations to calculate the number of instructions of the algorithm for a given input size (from which I could calculate the time required to calculate on a machine with a given speed). However, for Shor's algorithm, the most I can find is its complexity: O( (log N)$^3$ )}'', which receives 4,718 views. 
%``I'm a fledgling computer science scholar, and I'm being asked to write a paper which involves integer factorization. As a result, I'm having to look into Shor's algorithm on quantum computers. For the other algorithms, I was able to find specific equations to calculate the number of instructions of the algorithm for a given input size (from which I could calculate the time required to calculate on a machine with a given speed). However, for Shor's algorithm, the most I can find is its complexity: O( (log N)$^3$ ).'' and has 4,718 views.Also, 
This topic is linked to the questions category \texttt{theoretical} derived from RQ1.
This topic indicates developers' challenge in understanding the complexity of quantum algorithms.

\noindent\textbf{As quantum programming is oriented to searching solutions in a probabilistic space, which is counter-intuitive from the classical computing perspective, understanding \texttt{quantum execution results} is particularly challenging for developers.} 
As a Qubit can be 0 or 1 with a certain probability, a quantum program that has Qubits as its basic units can have many different states at the same time. 
The results of a quantum program are certain only when the results are observed (or ``measured''). 
Therefore, it is more challenging for developers to understand the results of quantum programs than that of classical programs. For example ``How to plot histogram or Bloch sphere for multiple circuits? I have tried to plot a histogram for the multiple circuits for the code given below. I think I have done it correctly but don't know why it's not working. Please help me out. If possible please solve it for the Bloch sphere'' %\footnote{https://quantumcomputing.stackexchange.com/questions/9224/how-to-plot-histogram-or-bloch-sphere-for-multiple-circuits}
%\heng{Can we have an example question here? I remember there are some examples.}
Future efforts are needed to interpret quantum program outputs. 

\input{tables/popularityQA}

\begin{figure}[!t]
\vspace{-2mm}
\centering

\centerline{\includegraphics[width=0.5\textwidth]{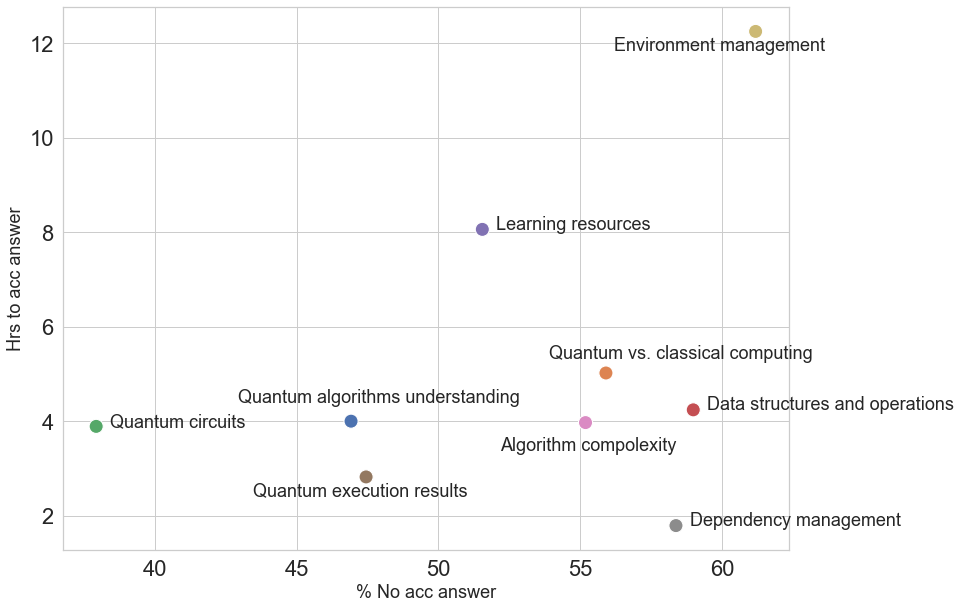}}
\caption{The difficulty aspect of QSE-related topics on Stack Exchange forums}
%\heng{make text further bigger} \heng{don't forget to change quantum backends to quantum execution results}
\label{figure:QAdifficulty}
\vspace{-8mm}
\end{figure}

\noindent\textbf{Posts related to \texttt{quantum vs.} \texttt{classical computing} are gaining the most attention from developers.} Since quantum computing is based on a new philosophy different from classical computing, developers often ask questions about the differences and look to understand the new doors quantum computing is opening. According to Table~\ref{table:QA popularity}, posts with this topic receive the highest median number of views.
%\texttt{quantum vs. classical computing} appears to be the least answered topic. This raises that it is difficult for the software engineer to answer a question about the difference between classical computing and quantum computing or quantum computing properties and applications. 
This may show that software engineers are eager to contribute to QSE by starting from the differences between the two paradigms.
However, as shown in Figure~\ref{figure:QAdifficulty}, posts on this topic are least likely to receive accepted answers. 
%but there is a limitation in the resources and works in QSE.
%\heng{talk about this topic and move the later discussions about this topic here.}
Our results indicate the need for resources and tools for bridging the knowledge gap between quantum computing and classical computing.
%In table \ref{table:QA popularity} we illustrate the result of the topic's popularity as defined previously then sorted by the median number of view. The popular topic is the one with more number of view, the best median score, and the most marked as a favorite\cite{9240667}. Form table \ref{table:QA popularity}, Quantum vs classical computing topic has the highest median number of view, score, and the number of favorites. We observe that dependency management is the least popular of the other 7 topics. This is because dependency management is a new topic that started to gain popularity in 2017 with the release of new quantum frameworks like Qiskit \cite{Qiskit}.

\textbf{Questions of some topics (e.g., \texttt{environment management}) are much more difficult than others to receive accepted answers}.
According to Figure \ref{figure:QAdifficulty}, the topic \texttt{environment management} is the most difficult topic to answer, with 61\% of posts not receiving an accepted answer and a median time of 12 hours to receive an accepted answer. 
\texttt{Learning resources}, \texttt{quantum vs. classical computing} and \texttt{data structures and operations} are also among the most difficult topics in terms of the ratio of posts getting accepted answers and the time to get one.
The results indicate the lack of community support in aspects such as setting up a development environment, searching for learning resources, and understanding differences between quantum computing and classical computing, which could impair the advancement of quantum software development practices.
%In contrast \textbf{\texttt{Dependency management}} is the least difficult topics, he has the lowest percentage of non-accepted answers (29\%) and the takes the least time to receive an answer (1.8 hours).

%in average developers need to wait around 210.32 hours and to only got 47.39\% of their questions answered. This show the difficulty of QSE in general and software engineer is struggling to get simplified documentation about quantum computing. Most of the provided resources discussing quantum physics which requires certain knowledge in this field. Unfortunately, this is not the case for software engineers. 

\begin{tcolorbox}
\vspace{-2mm}
From Q\&A forums, we derived nine topics discussed in QSE-related posts, including traditional software engineering topics (e.g., \texttt{environment management}) % and \texttt{dependency management}
and QSE-specific ones (e.g., \texttt{quantum execution results}). 
%and \texttt{quantum vs. classical computing}).
We highlighted some particularly challenging areas for QSE, such as interpreting quantum program outputs and bridging the knowledge gap between quantum computing and classical computing.
%Some topics (e.g., \texttt{Quantum simulator}) are gaining the highest attention while being much more difficult than others to find accepted answers.
\vspace{-2mm}
\end{tcolorbox}
%\begin{tcolorbox}
%\textbf{Summary finding (3):}Quantum Simulator, Environment management are among the most difficult topics to answer. In general, it appears that all topics are difficult to answer where Dependency management with 37.93\% not accepted answer takes on average 4-5 days to receive an answer.
%\end{tcolorbox}

%% file: tables/tableQAtopics.tex
\begin{table*}[htbp]
\vspace{-2mm}
\caption{Topics extracted from QSE related posts on Stack Exchange forums}
\centering
\begin{tabular}{llll}
\rowcolor[HTML]{EFEFEF} 
\hline
\multicolumn{1}{c}{\textbf{Topic (manual label)}} &
  \multicolumn{1}{c}{\textbf{Keywords}} &
  \multicolumn{1}{c}{\textbf{Description}} &
  \multicolumn{1}{c}{\textbf{\% Freq}} \\ \hline
Environment management &
  \begin{tabular}[c]{@{}l@{}}quot, error, python, build, code\end{tabular} &
  \begin{tabular}[c]{@{}l@{}} Development environment and build problems\end{tabular} &
  15.03 \\ \hline
Dependency management &
  \begin{tabular}[c]{@{}l@{}}qiskit, import, ibmq, operator, provider\end{tabular} &
  \begin{tabular}[c]{@{}l@{}}Library installation, use, and versioning issues\end{tabular} &
  14.82 \\ \hline
Algorithm complexity &
  \begin{tabular}[c]{@{}l@{}}time, problem, algorithm, number, function\end{tabular} &
  \begin{tabular}[c]{@{}l@{}}Quantum algorithm complexity and optimization\end{tabular} &
  14.06 \\ \hline
Quantum execution results &
  \begin{tabular}[c]{@{}l@{}}circuit, result, back-end, simulator, measure\end{tabular} &
  \begin{tabular}[c]{@{}l@{}}Quantum program execution results on quantum\\backends (e.g., simulators) \end{tabular} &
  13.22 \\ \hline
Learning resources &
  \begin{tabular}[c]{@{}l@{}}question, paper, work, understand, answer\end{tabular} &
  \begin{tabular}[c]{@{}l@{}}Searching for learning resources such as research\\ papers and tutorials \\ \end{tabular} &
  9.05 \\ \hline
Data structures and operations &
  \begin{tabular}[c]{@{}l@{}}matrix, return, array, datum, list\end{tabular} &
  \begin{tabular}[c]{@{}l@{}}Data structures (e.g., matrix, arrays and list) and\\ their operations in quantum programs \end{tabular} &
  8.81 \\ \hline
Quantum circuits &
  \begin{tabular}[c]{@{}l@{}}qubit, gate, control, operation, cirq\end{tabular} &
  \begin{tabular}[c]{@{}l@{}}Elements of quantum circuits (e.g., Qubits, gates)\\ and their operations\end{tabular} &
  8.66 \\ \hline
Quantum vs. classical computing &
  \begin{tabular}[c]{@{}l@{}}quantum, computer, classical, computing,\\ algorithm\end{tabular} &
  \begin{tabular}[c]{@{}l@{}}Comparisons between quantum and classical computing\\ or migrating classic algorithms to quantum computing\end{tabular} &
  8.30 \\ \hline
Quantum algorithms understanding &
  state, rangle, frac, theta, sqrt &
  \begin{tabular}[c]{@{}l@{}}Quantum algorithm explanation and interpretation\end{tabular} &
  7.51 \\ \hline
\end{tabular}
\label{table:QA_topic}
\vspace{-3mm}
\end{table*}

%% file: tables/popularityQA.tex
\begin{table}[!t]
\centering
%\caption{Quantum software engineering topics popularity}
\caption{Popularity of QSE-related topics on Stack Exchange forums}
\vspace{-5pt}
\setlength{\tabcolsep}{4.5pt}
\begin{tabular}{lrrr}
\rowcolor[HTML]{EFEFEF} 
\toprule
                   Topic name &   $\tilde{View}$ &   $\tilde{Score}$ & $\tilde{Favorite}$ \\
\midrule
  Quantum vs. classical computing &      147.5 &      3 &            1.5 \\
                 Quantum circuits &      107.0 &      2 &            1.0 \\
           Environment management &      106.0 &      1 &            1.0 \\
               Learning resources &      102.0 &      2 &            1.0 \\
                 Quantum execution results &       98.0 &      1 &            1.0 \\
 Quantum algorithms understanding &       97.5 &      2 &            1.0 \\
            Dependency management &       93.0 &      2 &            1.0 \\
            Algorithm compolexity &       87.5 &      1 &            1.0 \\
   Data structures and operations &       82.0 &      1 &            1.0 \\
\bottomrule
\end{tabular}
\label{table:QA popularity}
\vspace{-5mm}
\end{table}

%% file: Sections/RQ3.tex
\vspace{-5pt}
\subsection*{\textbf{RQ3: What QSE topics are raised in the issue reports of quantum-computing projects?}}

\input{tables/git_topics}

\setcounter{subsubsection}{0}
\subsubsection{\textbf{Motivation}}
Issue reports of quantum computing projects record developers' concerns and discussions when adding features or resolving issues in these projects.
The textual information in the issue reports may communicate developers' challenges when developing quantum computing applications. 
Therefore, we analyze the topics in the issue reports to understand the challenges developers are facing as well as the prevalence of these challenges.
While the questions on technical forums can provide information about developers' general challenges, the issue reports may communicate developers' challenges for specific problems (i.e., issues).

\subsubsection{\textbf{Approach}}

\noindent \textbf{Topic assignment and frequency.}
Our topic model on the issue report data generated 17 topics.
We follow the same process as described in RQ2 to manually assign meaningful labels to the automated topics, based on the top words in the topics and the content of the associated issue reports. 
During the manual assignment process, we found that some topics are similar to each other even though such similarity is not detected by the probabilistic topic model.
Therefore, we follow prior work~\cite{noei2019too, rosen2016mobile} and merged similar topics.
We also discarded one topic as we could not derive a meaningful label from the top words and the associated issue reports. 
In the end, we obtained 13 meaningful topics.
For each topic, we measure the percentage of the issue reports (i.e., frequency) that have it as the dominant topic (i.e., with the highest probability).
We also measure the number and percentage of the studied projects that have at least one issue report of each topic.

\noindent \textbf{Topic difficulty.}
To further understand developers' challenges in developing quantum computing applications, we measure the ``difficulty'' of the issue reports associated with each topic. 
As we cannot directly measure the ``difficulty'' of issue reports, we measure three indirect metrics for each topic: (1) the percentage of issue reports associated with the topic that is \texttt{closed}, (2) the median time required to close an issue (since its creation) associated with the topic, and (3) the median number of comments in an associated topic (intuitively, an issue report with more comments may be more difficult \cite{Khomh11}). 
For each topic, the associated issue reports refer to the issue reports that have it as the dominant topic.
\begin{comment}
\begin{itemize}
    \item We group the issues by dominant topic label
    \item We compute the difference between the closer date and the creation date in hour per issue. 
    \item We aggregate the time required by the issue to become closed and the issue number of comments with median aggregation function.
\end{itemize}
Hence \textbf{The topic is difficult if after a long waiting time he is closed.}. 
\end{comment}

\subsubsection{\textbf{Results}}
\begin{comment}
\heng{Three main messages: 
1) We derived 13 topics from GitHub issue reports, bringing new perspectives on developers' faced challenges related to QSE.
2) All the derived topics are general among the quantum computing projects, as each topic is present in 66\% to 90\% of the projects' issue reports.
3) Some topics are particularly challenging for developers, such as \texttt{data structures and operations}, \texttt{quantum circuits}, and \texttt{quantum execution results}.
}
\end{comment}

\noindent \textbf{We derived 13 topics from GitHub issue reports, bringing new perspectives to the challenges faced by QSE developers.
}
Table~\ref{table:git} shows the list of our derived topics, their descriptions, their percentage frequency in the studied issue reports, and the number of projects that have at least one issue report of the topic. 
Among the 13 topics, 6 of them (\texttt{learning resources}, \texttt{environment management}, \texttt{quantum circuits}, \texttt{quantum execution results}, \texttt{data structures and operations}, and \texttt{dependency management}) are overlapping with the topics derived from Stack Exchange posts (RQ2), and another 2 of them (\texttt{API change} and \texttt{API usage}) are overlapping with the categories of Stack Exchange questions derived in RQ1. 
This result indicates that the QSE-related challenges that we derived from forum posts indeed impact practical quantum program development in GitHub projects. 

Among the other five topics, two of them (i.e., \texttt{machine learning} and \texttt{quantum chemistry}) are related to the most popular and promising quantum computing application areas: machine learning and chemistry. %\heng{$\leftarrow$maybe add two citations one for each} 
%For example, an issues for chemistry topic `` Geometry Optimization - handle situation where calculation doesn't converge within maximum cycles Should probably just terminate geometry optimizations when the method doesn't converge within max\_cycle.
For example, an issue report associated with \texttt{quantum chemistry} raises an issue when using a molecular optimizer Python library: ``\textit{it leads to PyBerny optimizing an unconverged ground state energy, which generally leads to the geometry optimization never converging}''. %\footnote{https://github.com/pyscf/pyscf/issues/148} and 
%for machine learning topic `` Unable to use generator prediction I train a model and save it under the folder, then load the model to predict, but I get the error. ''
The other three topics (i.e., \texttt{quantum execution errors}, \texttt{unit testing}, \texttt{algorithm optimization}) are related to applications of traditional software engineering processes in quantum program development.

\noindent\textbf{All derived topics are general among the quantum computing projects, as each topic is present in the issue reports of 58\% to 90\% of the projects}. 
We observe that \texttt{learning resources} and \texttt{environment management} are the two most frequent topics and appear in 90\% and 88\% of all the studied projects, respectively, which once again highlights the need of efforts for developing learning resources and supporting developers in setting up their quantum program development environment.

\noindent\textbf{Some topics are particularly challenging for developers, such as \texttt{data structures} \texttt{and operations}, \texttt{quantum circuits}, and \texttt{quantum execution results}.}
Table \ref{table:gitDiff} shows the median time it takes to close an issue report and the number of comments in an issue report associated with each topic.
All the issues are closed at the time when we analyzed their status. 
The issues associated with each topic only have a median of one to two comments, indicating that developers' interactions on these issue reports are not intense.
\texttt{Data structures and operations} is also among the most difficult topics on forum posts (as discussed in RQ2).
However, the \texttt{Quantum circuits} and \texttt{quantum execution results} topics are not among the most difficult topics on forum posts, while they are two of the most difficult ones on GitHub issues, which indicates that quantum circuit issues and the interpretation of quantum program execution results are more difficult in specific problem contexts.
\input{tables/Git_difficulty}

\begin{tcolorbox}
\vspace{-2mm}
QSE-related challenges that we derived from forum posts indeed impact practical quantum program development in GitHub projects, while GitHub issues bring new perspectives on developers' faced challenges (e.g., on specific quantum computing applications such as machine learning).
In particular, we observe that the challenges are generally among the quantum computing projects. 
\vspace{-2mm}
\end{tcolorbox}

%% file: tables/git_topics.tex
\begin{table*}[!t]
\vspace{-8mm}
\centering
\caption{QSE-related topics derived from issue reports of quantum computing projects on GitHub} \
\resizebox{\textwidth}{!}{%
\begin{tabular}{@{}lllll@{}}
\rowcolor[HTML]{EFEFEF} 
\toprule
\multicolumn{1}{c}{\textbf{Manual label}} &
  \multicolumn{1}{c}{\textbf{Keywords}} &
  \multicolumn{1}{c}{\textbf{Description}} &
  \multicolumn{1}{c}{\textbf{\% Freq}} &
  \multicolumn{1}{c}{\textbf{\# Projects}}\\ 
  \midrule
Learning resources &
  \begin{tabular}[c]{@{}l@{}}summary, remove, tutorial, link, documentation\end{tabular} &
  \begin{tabular}[c]{@{}l@{}}Search for documentation, tutorials, websites, etc.\end{tabular} &
  14.94  & 109 (90.08\%)\\ \midrule
Environment management &
  \begin{tabular}[c]{@{}l@{}}build, include, library, release, variable\end{tabular} &
  \begin{tabular}[c]{@{}l@{}}Development environment and build problems\end{tabular} &
  13.72 & 107 (88.43\%) \\ \midrule
API change &
  \begin{tabular}[c]{@{}l@{}}version, qiskit, code, issue, update\end{tabular} &
  \begin{tabular}[c]{@{}l@{}}API update or deprecation issues\end{tabular} &
  11.65 & 98 (80.99\%) \\ \midrule
Quantum circuits &
  \begin{tabular}[c]{@{}l@{}}gate, circuit, qubit, operation, control\end{tabular} &
  \begin{tabular}[c]{@{}l@{}}Elements of quantum circuits (e.g., Qubits, gates) 
  \\and their operations\end{tabular} &
  8.30 & 70 (57.85\%) \\ \midrule
Quantum chemistry &
  \begin{tabular}[c]{@{}l@{}}input, calculation, basis, energy, pyscf\end{tabular} &
   \begin{tabular}[c]{@{}l@{}}Issues with quantum chemistry libraries (e.g., PySCF)\end{tabular} &
  6.72 & 76 (62.80\%)\\ \midrule
Quantum execution errors &
  \begin{tabular}[c]{@{}l@{}}error, artiq, follow, experiment, device\end{tabular} &
  \begin{tabular}[c]{@{}l@{}}Errors in the execution of quantum programs\end{tabular} &
  6.38 & 80 (66.12\%)\\ \midrule
Unit testing &
  \begin{tabular}[c]{@{}l@{}}test, check,   fail, unit, script\end{tabular} &
  \begin{tabular}[c]{@{}l@{}}Unit  testing failures\end{tabular} &
  6.32 & 87 (71.90\%)\\ \midrule
API usage &
  \begin{tabular}[c]{@{}l@{}}function, method, class, parameter, call\end{tabular} &
  \begin{tabular}[c]{@{}l@{}}How to use an API \end{tabular} &
  5.81 & 80 (66.11\%)\\ \midrule
Quantum execution results &
  \begin{tabular}[c]{@{}l@{}}state, number, result, time, measurement\end{tabular} &
  \begin{tabular}[c]{@{}l@{}}Quantum program execution results (i.e., measured state)\end{tabular} &
  5.76 & 89 (73.55\%) \\ \midrule
Data structures and operations &
  \begin{tabular}[c]{@{}l@{}}implement, operator, matrix, problem, array\end{tabular} &
  \begin{tabular}[c]{@{}l@{}}Data structures (e.g., matrix, arrays and list) \\and their operations \end{tabular} &
  5.73 & 88 (72.73\%)\\ \midrule
Machine learning &
  \begin{tabular}[c]{@{}l@{}}model, datum, dataset, layer, benchmark\end{tabular} &
  \begin{tabular}[c]{@{}l@{}}Quantum computing application in machine learning\end{tabular} &
  5.26 & 75 (61.9\%) \\ \midrule
Dependency management &
  \begin{tabular}[c]{@{}l@{}}file, python, import, package, install \end{tabular} &
  \begin{tabular}[c]{@{}l@{}}Library installation, use and versioning issues \end{tabular} &
  5.23 & 93 (76.86\%) \\ \midrule
Algorithm optimization &
  \begin{tabular}[c]{@{}l@{}}case, time, optimization, long, performance\end{tabular} &
  \begin{tabular}[c]{@{}l@{}}Program performance and algorithm optimization \end{tabular} &
  4.19 & 83 (68.6\%) \\ \bottomrule
\end{tabular}
}
\label{table:git}
\vspace{-5mm}
\end{table*}

%% file: tables/Git_difficulty.tex
\begin{table}[!t]
\centering
%\caption{QSE Topics difficulty}
\vspace{-5pt}
\caption{The difficulty aspect of QSE-related topics on GitHub issues}
\vspace{-4pt}
\begin{tabular}{llrr}
\rowcolor[HTML]{EFEFEF} 
\toprule
                   Topic name &     $\tilde{Hr\,\, to\,\, close}$ &  $\tilde{\#\, comments}$ \\
\midrule
   Data structures and operations &  151.40 &         1 \\
                 Quantum circuits &  114.98 &         1 \\
               Quantum execution results &   98.02 &         1 \\
                 Machine learning &   94.68 &         2 \\
                        API usage &   80.70 &         1 \\
                        Quantum chemistry &   62.89 &         2 \\
    Quantum execution errors &   59.44 &         2 \\
                       API change &   47.34 &         1 \\
                     Algorithm optimization &   39.83 &         1 \\
            Dependency management &   32.40 &         2 \\
                     Unit testing &   28.26 &         1 \\
               Learning resources &   27.02 &         1 \\
          Environment management &   21.57 &         1 \\
\bottomrule
\multicolumn{3}{l}{All the issues are closed at the time we analyzed their status.}
\end{tabular}
\label{table:gitDiff}
\vspace{-7.5mm}
\end{table}

%% file: Sections/Threats.tex
\vspace{-4pt}
\section{Threats to Validity} \label{sec:threats}

\noindent \textbf{External validity.}
In this work, we analyze four Stack Exchange forums and 122 GitHub repositories to understand the challenges of QSE. Our studied forum posts and GitHub issues may not cover all the ones that are related to QSE. Developers may also communicate their discussions in other media (e.g., mailing lists). Future work considering other data sources may complement our study.
In addition, we identify and collect the posts from Q\&A forums using a selected set of tags. Our analysis may miss some QSE tags. However, to alleviate this threat, we follow prior work~\cite{uddin2021understanding, 9240667} and use an iterative method to identify the relevant tags.

\noindent \textbf{Internal validity.}
In this work, we use topic models to cluster the forum posts and GitHub issue reports, based on the intuition that the same clusters would have similar textual information. However, different clusters of posts and issue reports may exist when a different approach is used. 
To ensure the quality of the clusters, we manually reviewed the resulting topics, merged similar topics when needed, and assigned meaningful labels to them.

\noindent\textbf{Construct validity.}
In RQ1, we manually analyze the categories of QSE-related questions on technical Q\&A forums. Our results may be subjective and depend on the judgment of the researchers who conducted the manual analysis. To mitigate the bias, two authors of the paper collectively conducted an manual analysis and reached a substantial agreement, indicating the reliability of the analysis results. A third author also participated in the discussions during the manual analysis, to ensure the quality of the results. 
In RQ2 and RQ3, the parameters of the topic models (e.g., the number of topics $K$) may impact our findings. To mitigate this threat, following previous work ~\cite{uddin2021understanding} \cite{9240667}, we did multiple experiments and use the topic coherence score to select the most suitable parameters.
The manual labeling of topics can be subjective. To reduce this threat, the authors read each topic's top 20 keywords and the top 15 highest contributed posts to the topic. We followed a clear-cut approach adapted in previous works~\cite{uddin2021understanding} \cite{9240667}. In addition, in our analysis of the QSE posts, we did not filter posts using the number of comments, votes or answers (as done in prior work~\cite{farhana2019synthesizing}), which may lead to noise in the analyzed posts (e.g., low-quality posts). 
We made this decision since QSE is a new topic and the number of posts in the Q\&A forums is relatively small. 
%this will likely result to noise, to reduce this threat, the first author manually inspected a sample of the posts.

%\heng{always need punctuation!} 
%\Foutse{we described a process in which three people at least were involved...we need to mention it here as well}

%\Foutse{can we provide a replication package? to cover reliability validity threats?}
% Maybe to add: determining the parameters of LDA

%% file: Sections/Conclusions.tex
\section{Conclusions} \label{sec:conclusions}

%Quantum software engineering (QSE) is an emerging area that is attracting more and more attention from the research community. 
%In this work, we study the 
This paper examines challenges quantum program developers are facing by analyzing Stack Exchange forums posts related to QSE and the  %where developers ask QSE-related questions and 
GitHub issue reports of  %where developers raise issues in practical 
quantum computing projects. 
Results indicate that QSE developers face traditional software engineering challenges (e.g., dependency management) as well as QSE-specific challenges (e.g., interpreting quantum program execution results). 
In particular, some QSE-related areas (e.g., bridging the knowledge gap between quantum and classical computing) are gaining the highest attention from developers while being very %the most 
challenging to them. 
%We observe that these challenges are usually general among practical quantum computing projects.
%In particular, some QSE-related topics (e.g., quantum vs. classical computing) %\heng{to update} 
%are gaining the highest attention while being the most challenging to practitioners.
As the initial effort for understanding QSE-related challenges perceived by developers, our work shed light on future opportunities in QSE (e.g., supporting explanations of theory behind quantum program code and the interpretations of quantum program execution results).

\begin{comment}
Based on an existing taxonomy of question categories on Stack Overflow, we first perform a qualitative analysis of the categories of QSE-related questions. We find nine categories of questions, among which four are new or more frequent in QSE-related posts.
We also use automated topic modeling to uncover the topics in QSE-related Q\&A posts and issue reports. We find overlapping and mutually complementary topics in these two types of data, which include traditional software engineering topics (e.g., dependency management) and QSE-specific topics (e.g., quantum program execution). 
In particular, some topics (e.g., Quantum simulator) are gaining the highest attention while being much more difficult than others to find accepted answers.
\end{comment}

%% file: quantumcomp.bbl
\begin{thebibliography}{10}
\providecommand{\url}[1]{#1}
\csname url@samestyle\endcsname
\providecommand{\newblock}{\relax}
\providecommand{\bibinfo}[2]{#2}
\providecommand{\BIBentrySTDinterwordspacing}{\spaceskip=0pt\relax}
\providecommand{\BIBentryALTinterwordstretchfactor}{4}
\providecommand{\BIBentryALTinterwordspacing}{\spaceskip=\fontdimen2\font plus
\BIBentryALTinterwordstretchfactor\fontdimen3\font minus
  \fontdimen4\font\relax}
\providecommand{\BIBforeignlanguage}[2]{{%
\expandafter\ifx\csname l@#1\endcsname\relax
\typeout{** WARNING: IEEEtran.bst: No hyphenation pattern has been}%
\typeout{** loaded for the language `#1'. Using the pattern for}%
\typeout{** the default language instead.}%
\else
\language=\csname l@#1\endcsname
\fi
#2}}
\providecommand{\BIBdecl}{\relax}
\BIBdecl

\bibitem{knight2018serious}
W.~Knight, ``Serious quantum computers are finally here. what are we going to
  do with them,'' \emph{MIT Technology Review}, vol.~30, 2018.

\bibitem{maslov2018outlook}
D.~Maslov, Y.~Nam, and J.~Kim, ``An outlook for quantum computing [point of
  view],'' \emph{Proceedings of the IEEE}, vol. 107, no.~1, pp. 5--10, 2018.

\bibitem{zhao2020quantum}
J.~Zhao, ``Quantum software engineering: Landscapes and horizons,'' \emph{arXiv
  preprint arXiv:2007.07047}, 2020.

\bibitem{ibm}
IBM, ``{IBM Quantum Computing},'' \url{https://www.ibm.com/quantum-computing/},
  {Last accessed 05/04/2021}.

\bibitem{dirac1981principles}
P.~A.~M. Dirac, \emph{The principles of quantum mechanics}.\hskip 1em plus
  0.5em minus 0.4em\relax Oxford university press, 1981, no.~27.

\bibitem{schrodinger1935discussion}
E.~Schr{\"o}dinger, ``Discussion of probability relations between separated
  systems,'' in \emph{Mathematical Proceedings of the Cambridge Philosophical
  Society}, vol.~31, no.~4.\hskip 1em plus 0.5em minus 0.4em\relax Cambridge
  University Press, 1935, pp. 555--563.

\bibitem{mueck2017quantum}
L.~Mueck, ``Quantum software,'' \emph{Nature}, vol. 549, no. 171, 2017.

\bibitem{piattini2020talavera}
M.~Piattini, G.~Peterssen, R.~P{\'e}rez-Castillo, J.~L. Hevia, M.~A. Serrano,
  G.~Hern{\'a}ndez, I.~G.~R. de~Guzm{\'a}n, C.~A. Paradela, M.~Polo, E.~Murina
  \emph{et~al.}, ``The talavera manifesto for quantum software engineering and
  programming.'' in \emph{The First International Workshop on the Quantum
  Software Engineering \& Programming}, ser. QANSWER '20, 2020, pp. 1--5.

\bibitem{omer2003qcl}
B.~{\"O}mer, ``Qcl-a programming language for quantum computers,''
  \emph{Software available on-line at http://tph. tuwien. ac. at/\~{}
  oemer/qcl. html}, 2003.

\bibitem{Qiskit}
H.~Abraham, AduOffei, R.~Agarwal, I.~Y. Akhalwaya, and et~al., ``Qiskit: An
  open-source framework for quantum computing,'' 2019.

\bibitem{quantumai}
Google, ``{Google QuantumAI},'' \url{https://quantumai.google}, {Last accessed
  05/04/2021}.

\bibitem{microsoft.azure}
Microsoft, ``{Microsoft Azure Quantum Service},''
  \url{https://azure.microsoft.com/en-ca/services/quantum/}, {Last accessed
  05/04/2021}.

\bibitem{biamonte2017quantum}
J.~Biamonte, P.~Wittek, N.~Pancotti, P.~Rebentrost, N.~Wiebe, and S.~Lloyd,
  ``Quantum machine learning,'' \emph{Nature}, vol. 549, no. 7671, pp.
  195--202, 2017.

\bibitem{guerreschi2017practical}
G.~G. Guerreschi and M.~Smelyanskiy, ``Practical optimization for hybrid
  quantum-classical algorithms,'' \emph{arXiv preprint arXiv:1701.01450}, 2017.

\bibitem{mailloux2016post}
L.~O. Mailloux, C.~D. Lewis~II, C.~Riggs, and M.~R. Grimaila, ``Post-quantum
  cryptography: what advancements in quantum computing mean for it
  professionals,'' \emph{IT Professional}, vol.~18, no.~5, pp. 42--47, 2016.

\bibitem{reiher2017elucidating}
M.~Reiher, N.~Wiebe, K.~M. Svore, D.~Wecker, and M.~Troyer, ``Elucidating
  reaction mechanisms on quantum computers,'' \emph{Proceedings of the National
  Academy of Sciences}, vol. 114, no.~29, pp. 7555--7560, 2017.

\bibitem{piattini2020quantum}
M.~Piattini, G.~Peterssen, and R.~P{\'e}rez-Castillo, ``Quantum computing: A
  new software engineering golden age,'' \emph{ACM SIGSOFT Software Engineering
  Notes}, vol.~45, no.~3, pp. 12--14, 2020.

\bibitem{moguel2020roadmap}
E.~Moguel, J.~Berrocal, J.~Garc{\'\i}a-Alonso, and J.~M. Murillo, ``A roadmap
  for quantum software engineering: applying the lessons learned from the
  classics,'' in \emph{Proceedings of the 1st International Workshop on
  Software Engineering ++\& Technology}, ser. Q-SET '20, 2020.

\bibitem{barbosa2020software}
L.~S. Barbosa, ``Software engineering for'quantum advantage','' in
  \emph{Proceedings of the IEEE/ACM 42nd International Conference on Software
  Engineering Workshops}, 2020, pp. 427--429.

\bibitem{piattini2021toward}
M.~Piattini, M.~Serrano, R.~Perez-Castillo, G.~Petersen, and J.~L. Hevia,
  ``Toward a quantum software engineering,'' \emph{IT Professional}, vol.~23,
  no.~1, pp. 62--66, 2021.

\bibitem{Beyer2021WhatKO}
S.~Beyer, C.~Macho, M.~D. Penta, and M.~Pinzger, ``What kind of questions do
  developers ask on stack overflow? a comparison of automated approaches to
  classify posts into question categories,'' in \emph{Software Engineering},
  2021.

\bibitem{kaye2007introduction}
P.~Kaye, R.~Laflamme, M.~Mosca \emph{et~al.}, \emph{An introduction to quantum
  computing}.\hskip 1em plus 0.5em minus 0.4em\relax Oxford University Press on
  Demand, 2007.

\bibitem{DBLP:journals/corr/abs-1803-07407}
\BIBentryALTinterwordspacing
B.~Sodhi, ``Quality attributes on quantum computing platforms,'' \emph{CoRR},
  vol. abs/1803.07407, 2018. [Online]. Available:
  \url{http://arxiv.org/abs/1803.07407}
\BIBentrySTDinterwordspacing

\bibitem{garhwal2019quantum}
S.~Garhwal, M.~Ghorani, and A.~Ahmad, ``Quantum programming language: A
  systematic review of research topic and top cited languages,'' \emph{Archives
  of Computational Methods in Engineering}, pp. 1--22, 2019.

\bibitem{doi:10.1098/rspa.1985.0070}
\BIBentryALTinterwordspacing
D.~Deutsch and R.~Penrose, ``Quantum theory, the church\&\#x2013;turing
  principle and the universal quantum computer,'' \emph{Proceedings of the
  Royal Society of London. A. Mathematical and Physical Sciences}, vol. 400,
  no. 1818, pp. 97--117, 1985. [Online]. Available:
  \url{https://royalsocietypublishing.org/doi/abs/10.1098/rspa.1985.0070}
\BIBentrySTDinterwordspacing

\bibitem{osti_366453}
\BIBentryALTinterwordspacing
E.~Knill, ``Conventions for quantum pseudocode,'' 6 1996. [Online]. Available:
  \url{https://www.osti.gov/biblio/366453}
\BIBentrySTDinterwordspacing

\bibitem{SandersJW:quap}
J.~W. Sanders and P.~Zuliani, ``Quantum programming,'' in \emph{Mathematics of
  Program Construction}, ser. Lecture Notes in Computer Science, vol.
  1837.\hskip 1em plus 0.5em minus 0.4em\relax Springer, 2000, pp. 80--99.

\bibitem{doi:10.1142/S0219749908004031}
\BIBentryALTinterwordspacing
H.~MLNAŘÍK, ``Semantics of quantum programming language lanq,''
  \emph{International Journal of Quantum Information}, vol.~06, no. supp01, pp.
  733--738, 2008. [Online]. Available:
  \url{https://doi.org/10.1142/S0219749908004031}
\BIBentrySTDinterwordspacing

\bibitem{10.1145/3183895.3183901}
\BIBentryALTinterwordspacing
K.~Svore, A.~Geller, M.~Troyer, J.~Azariah, C.~Granade, B.~Heim,
  V.~Kliuchnikov, M.~Mykhailova, A.~Paz, and M.~Roetteler, ``Q\#: Enabling
  scalable quantum computing and development with a high-level dsl,'' in
  \emph{Proceedings of the Real World Domain Specific Languages Workshop 2018},
  ser. RWDSL2018.\hskip 1em plus 0.5em minus 0.4em\relax New York, NY, USA:
  Association for Computing Machinery, 2018. [Online]. Available:
  \url{https://doi.org/10.1145/3183895.3183901}
\BIBentrySTDinterwordspacing

\bibitem{larose2019overview}
R.~LaRose, ``Overview and comparison of gate level quantum software
  platforms,'' \emph{Quantum}, vol.~3, p. 130, 2019.

\bibitem{StackExchange.data}
B.~Vasilescu, ``Academic papers using stack exchange data,''
  \url{{https://meta.stackexchange.com/questions/134495/academic-papers-using-stack-exchange-data}},
  {Last accessed 05/04/2021}.

\bibitem{wang2013empirical}
S.~Wang, D.~Lo, and L.~Jiang, ``An empirical study on developer interactions in
  stackoverflow,'' in \emph{Proceedings of the 28th Annual ACM Symposium on
  Applied Computing}, 2013, pp. 1019--1024.

\bibitem{chen2019modeling}
H.~Chen, J.~Coogle, and K.~Damevski, ``Modeling stack overflow tags and topics
  as a hierarchy of concepts,'' \emph{Journal of Systems and Software}, vol.
  156, pp. 283--299, 2019.

\bibitem{allamanis2013and}
M.~Allamanis and C.~Sutton, ``Why, when, and what: analyzing stack overflow
  questions by topic, type, and code,'' in \emph{2013 10th Working Conference
  on Mining Software Repositories (MSR)}.\hskip 1em plus 0.5em minus
  0.4em\relax IEEE, 2013, pp. 53--56.

\bibitem{linares2013exploratory}
M.~Linares-V{\'a}squez, B.~Dit, and D.~Poshyvanyk, ``An exploratory analysis of
  mobile development issues using stack overflow,'' in \emph{2013 10th Working
  Conference on Mining Software Repositories (MSR)}.\hskip 1em plus 0.5em minus
  0.4em\relax IEEE, 2013, pp. 93--96.

\bibitem{rosen2016mobile}
C.~Rosen and E.~Shihab, ``What are mobile developers asking about? a large
  scale study using stack overflow,'' \emph{Empirical Software Engineering},
  vol.~21, no.~3, pp. 1192--1223, 2016.

\bibitem{venkatesh2016client}
P.~K. Venkatesh, S.~Wang, F.~Zhang, Y.~Zou, and A.~E. Hassan, ``What do client
  developers concern when using web apis? an empirical study on developer
  forums and stack overflow,'' in \emph{2016 IEEE International Conference on
  Web Services (ICWS)}.\hskip 1em plus 0.5em minus 0.4em\relax IEEE, 2016, pp.
  131--138.

\bibitem{alshangiti2019developing}
M.~Alshangiti, H.~Sapkota, P.~K. Murukannaiah, X.~Liu, and Q.~Yu, ``Why is
  developing machine learning applications challenging? a study on stack
  overflow posts,'' in \emph{2019 ACM/IEEE International Symposium on Empirical
  Software Engineering and Measurement (ESEM)}.\hskip 1em plus 0.5em minus
  0.4em\relax IEEE, 2019, pp. 1--11.

\bibitem{ahmed2018concurrency}
S.~Ahmed and M.~Bagherzadeh, ``What do concurrency developers ask about? a
  large-scale study using stack overflow,'' in \emph{Proceedings of the 12th
  ACM/IEEE International Symposium on Empirical Software Engineering and
  Measurement}, 2018, pp. 1--10.

\bibitem{yang2016security}
X.-L. Yang, D.~Lo, X.~Xia, Z.-Y. Wan, and J.-L. Sun, ``What security questions
  do developers ask? a large-scale study of stack overflow posts,''
  \emph{Journal of Computer Science and Technology}, vol.~31, no.~5, pp.
  910--924, 2016.

\bibitem{zou2017towards}
J.~Zou, L.~Xu, M.~Yang, X.~Zhang, and D.~Yang, ``Towards comprehending the
  non-functional requirements through developers’ eyes: An exploration of
  stack overflow using topic analysis,'' \emph{Information and Software
  Technology}, vol.~84, pp. 19--32, 2017.

\bibitem{zou2015non}
J.~Zou, L.~Xu, W.~Guo, M.~Yan, D.~Yang, and X.~Zhang, ``Which non-functional
  requirements do developers focus on? an empirical study on stack overflow
  using topic analysis,'' in \emph{2015 IEEE/ACM 12th Working Conference on
  Mining Software Repositories}.\hskip 1em plus 0.5em minus 0.4em\relax IEEE,
  2015, pp. 446--449.

\bibitem{zhang2015multi}
Y.~Zhang, D.~Lo, X.~Xia, and J.-L. Sun, ``Multi-factor duplicate question
  detection in stack overflow,'' \emph{Journal of Computer Science and
  Technology}, vol.~30, no.~5, pp. 981--997, 2015.

\bibitem{treude2019predicting}
C.~Treude and M.~Wagner, ``Predicting good configurations for github and stack
  overflow topic models,'' in \emph{2019 IEEE/ACM 16th International Conference
  on Mining Software Repositories (MSR)}.\hskip 1em plus 0.5em minus
  0.4em\relax IEEE, 2019, pp. 84--95.

\bibitem{zhang2014novel}
T.~Zhang, G.~Yang, B.~Lee, and E.~K. Lua, ``A novel developer ranking algorithm
  for automatic bug triage using topic model and developer relations,'' in
  \emph{2014 21st Asia-Pacific Software Engineering Conference}, vol.~1.\hskip
  1em plus 0.5em minus 0.4em\relax IEEE, 2014, pp. 223--230.

\bibitem{xia2013accurate}
X.~Xia, D.~Lo, X.~Wang, and B.~Zhou, ``Accurate developer recommendation for
  bug resolution,'' in \emph{2013 20th Working Conference on Reverse
  Engineering (WCRE)}.\hskip 1em plus 0.5em minus 0.4em\relax IEEE, 2013, pp.
  72--81.

\bibitem{naguib2013bug}
H.~Naguib, N.~Narayan, B.~Br{\"u}gge, and D.~Helal, ``Bug report assignee
  recommendation using activity profiles,'' in \emph{2013 10th Working
  Conference on Mining Software Repositories (MSR)}.\hskip 1em plus 0.5em minus
  0.4em\relax IEEE, 2013, pp. 22--30.

\bibitem{xia2016improving}
X.~Xia, D.~Lo, Y.~Ding, J.~M. Al-Kofahi, T.~N. Nguyen, and X.~Wang, ``Improving
  automated bug triaging with specialized topic model,'' \emph{IEEE
  Transactions on Software Engineering}, vol.~43, no.~3, pp. 272--297, 2016.

\bibitem{hindle2016contextual}
A.~Hindle, A.~Alipour, and E.~Stroulia, ``A contextual approach towards more
  accurate duplicate bug report detection and ranking,'' \emph{Empirical
  Software Engineering}, vol.~21, no.~2, pp. 368--410, 2016.

\bibitem{nguyen2012duplicate}
A.~T. Nguyen, T.~T. Nguyen, T.~N. Nguyen, D.~Lo, and C.~Sun, ``Duplicate bug
  report detection with a combination of information retrieval and topic
  modeling,'' in \emph{2012 Proceedings of the 27th IEEE/ACM International
  Conference on Automated Software Engineering}.\hskip 1em plus 0.5em minus
  0.4em\relax IEEE, 2012, pp. 70--79.

\bibitem{zou2016duplication}
J.~Zou, L.~Xu, M.~Yang, M.~Yan, D.~Yang, and X.~Zhang, ``Duplication detection
  for software bug reports based on topic model,'' in \emph{2016 9th
  International Conference on Service Science (ICSS)}.\hskip 1em plus 0.5em
  minus 0.4em\relax IEEE, 2016, pp. 60--65.

\bibitem{nguyen2011topic}
A.~T. Nguyen, T.~T. Nguyen, J.~Al-Kofahi, H.~V. Nguyen, and T.~N. Nguyen, ``A
  topic-based approach for narrowing the search space of buggy files from a bug
  report,'' in \emph{2011 26th IEEE/ACM International Conference on Automated
  Software Engineering (ASE 2011)}.\hskip 1em plus 0.5em minus 0.4em\relax
  IEEE, 2011, pp. 263--272.

\bibitem{martie2012trendy}
L.~Martie, V.~K. Palepu, H.~Sajnani, and C.~Lopes, ``Trendy bugs: Topic trends
  in the android bug reports,'' in \emph{2012 9th IEEE Working Conference on
  Mining Software Repositories (MSR)}.\hskip 1em plus 0.5em minus 0.4em\relax
  IEEE, 2012, pp. 120--123.

\bibitem{aggarwal2014mining}
A.~Aggarwal, G.~Waghmare, and A.~Sureka, ``Mining issue tracking systems using
  topic models for trend analysis, corpus exploration, and understanding
  evolution,'' in \emph{Proceedings of the 3rd International Workshop on
  Realizing Artificial Intelligence Synergies in Software Engineering}, 2014,
  pp. 52--58.

\bibitem{stackoverflow}
``{Stack Overflow forum},'' \url{{https://stackoverflow.com}}, {Last accessed
  05/04/2021}.

\bibitem{quantumcomputing.stackexchange}
``{Stack Exchange Quantum Computing forum},''
  \url{{https://quantumcomputing.stackexchange.com}}, {Last accessed
  05/04/2021}.

\bibitem{cs.stackexchange}
``{Stack Exchange Computer Science forum},''
  \url{{https://cs.stackexchange.com}}, {Last accessed 05/04/2021}.

\bibitem{ai.stackexchange}
``{Stack Exchange Artificial Intelligence forum},''
  \url{{https://ai.stackexchange.com}}, {Last accessed 05/04/2021}.

\bibitem{data.stackexchange}
``{}stack exchange data explorer.''

\bibitem{uddin2021understanding}
G.~Uddin, O.~Baysal, L.~Guerrouj, and F.~Khomh, ``Understanding how and why
  developers seek and analyze api-related opinions,'' 2021.

\bibitem{9240667}
M.~Openja, B.~Adams, and F.~Khomh, ``Analysis of modern release engineering
  topics : – a large-scale study using stackoverflow –,'' in \emph{2020
  IEEE International Conference on Software Maintenance and Evolution (ICSME)},
  2020, pp. 104--114.

\bibitem{10.1109/MSR.2017.5}
B.~Cartaxo, G.~Pinto, D.~Ribeiro, F.~K. Kamei, R.~Santos, F.~Silva, and
  S.~Soares, ``Using q\&a websites as a method for assessing systematic
  reviews,'' 05 2017.

\bibitem{Businge2018CloneBasedVM}
J.~Businge, M.~Openja, S.~Nadi, E.~Bainomugisha, and T.~Berger, ``Clone-based
  variability management in the android ecosystem,'' \emph{2018 IEEE
  International Conference on Software Maintenance and Evolution (ICSME)}, pp.
  625--634, 2018.

\bibitem{Businge2019StudyingAA}
J.~Businge, M.~Openja, D.~Kavaler, E.~Bainomugisha, F.~Khomh, and V.~Filkov,
  ``Studying android app popularity by cross-linking github and google play
  store,'' \emph{2019 IEEE 26th International Conference on Software Analysis,
  Evolution and Reengineering (SANER)}, pp. 287--297, 2019.

\bibitem{githubrestapi}
``{GitHub REST API},'' \url{https://developer.github.com/v3/}, {Last accessed
  05/04/2021}.

\bibitem{potterStemming}
P.~Willett, ``The porter stemming algorithm: Then and now,'' \emph{Program
  electronic library and information systems}, vol.~40, 07 2006.

\bibitem{PerformanceAnalysis:Stemming}
V.~Gurusamy and S.~Kannan, ``Performance analysis: Stemming algorithm for the
  english language,'' \emph{International Journal for Scientific Research and
  Development}, vol.~5, pp. 2321--613, 08 2017.

\bibitem{blei2003latent}
D.~M. Blei, A.~Y. Ng, and M.~I. Jordan, ``Latent dirichlet allocation,''
  \emph{the Journal of machine Learning research}, vol.~3, pp. 993--1022, 2003.

\bibitem{chen2016survey}
T.-H. Chen, S.~W. Thomas, and A.~E. Hassan, ``A survey on the use of topic
  models when mining software repositories,'' \emph{Empirical Software
  Engineering}, vol.~21, no.~5, pp. 1843--1919, 2016.

\bibitem{barua2014developers}
A.~Barua, S.~W. Thomas, and A.~E. Hassan, ``What are developers talking about?
  an analysis of topics and trends in stack overflow,'' \emph{Empirical
  Software Engineering}, vol.~19, no.~3, pp. 619--654, 2014.

\bibitem{inproceedings}
R.~Rossi and S.~Rezende, ``Generating features from textual documents through
  association rules,'' 01 2011.

\bibitem{McCallumMALLET}
A.~K. McCallum, ``Mallet: A machine learning for language toolkit,'' 2002,
  http://mallet.cs.umass.edu.

\bibitem{Rder2015ExploringTS}
M.~R{\"o}der, A.~Both, and A.~Hinneburg, ``Exploring the space of topic
  coherence measures,'' \emph{Proceedings of the Eighth ACM International
  Conference on Web Search and Data Mining}, 2015.

\bibitem{Gensim.CoherenceModel}
``Gensim coherencemodel implimentation,''
  \url{{https://radimrehurek.com/gensim/models/coherencemodel.html}}, {Last
  accessed 05/04/2021}.

\bibitem{Han2020WhatDP}
J.~Han, E.~Shihab, Z.~Wan, S.~Deng, and X.~Xia, ``What do programmers discuss
  about deep learning frameworks,'' \emph{Empirical Software Engineering},
  vol.~25, pp. 2694--2747, 2020.

\bibitem{beyer2020kind}
S.~Beyer, C.~Macho, M.~Di~Penta, and M.~Pinzger, ``What kind of questions do
  developers ask on stack overflow? a comparison of automated approaches to
  classify posts into question categories,'' \emph{Empirical Software
  Engineering}, vol.~25, no.~3, pp. 2258--2301, 2020.

\bibitem{li2020qualitative}
H.~Li, W.~Shang, B.~Adams, M.~Sayagh, and A.~E. Hassan, ``A qualitative study
  of the benefits and costs of logging from developers' perspectives,''
  \emph{IEEE Transactions on Software Engineering}, 2020.

\bibitem{MaryMcHugh}
M.~McHugh, ``Interrater reliability: The kappa statistic,'' \emph{Biochemia
  medica : časopis Hrvatskoga društva medicinskih biokemičara / HDMB},
  vol.~22, pp. 276--82, 10 2012.

\bibitem{cirq_developers_2021_4586899}
\BIBentryALTinterwordspacing
``Cirq,'' Mar. 2021, {See full list of authors on Github: https://github
  .com/quantumlib/Cirq/graphs/contributors}. [Online]. Available:
  \url{https://doi.org/10.5281/zenodo.4586899}
\BIBentrySTDinterwordspacing

\bibitem{Yang:Xin-Li:Lo}
X.-L. Yang, D.~Lo, X.~Xia, Z.~Wan, and J.-L. Sun, ``What security questions do
  developers ask? a large-scale study of stack overflow posts,'' \emph{Journal
  of Computer Science and Technology}, vol.~31, pp. 910--924, 09 2016.

\bibitem{10.1145/3338906.3338939}
\BIBentryALTinterwordspacing
M.~Bagherzadeh and R.~Khatchadourian, ``Going big: A large-scale study on what
  big data developers ask,'' in \emph{Proceedings of the 2019 27th ACM Joint
  Meeting on European Software Engineering Conference and Symposium on the
  Foundations of Software Engineering}, ser. ESEC/FSE 2019.\hskip 1em plus
  0.5em minus 0.4em\relax New York, NY, USA: Association for Computing
  Machinery, 2019, p. 432–442. [Online]. Available:
  \url{https://doi.org/10.1145/3338906.3338939}
\BIBentrySTDinterwordspacing

\bibitem{noei2019too}
E.~Noei, F.~Zhang, and Y.~Zou, ``Too many user-reviews, what should app
  developers look at first?'' \emph{IEEE Transactions on Software Engineering},
  2019.

\bibitem{Khomh11}
F.~Khomh, B.~Chan, Y.~Zou, and A.~E. Hassan, ``An entropy evaluation approach
  for triaging field crashes: A case study of mozilla firefox,'' in \emph{2011
  18th Working Conference on Reverse Engineering}, 2011, pp. 261--270.

\bibitem{farhana2019synthesizing}
E.~Farhana, N.~Imtiaz, and A.~Rahman, ``Synthesizing program execution time
  discrepancies in julia used for scientific software,'' in \emph{2019 IEEE
  International Conference on Software Maintenance and Evolution
  (ICSME)}.\hskip 1em plus 0.5em minus 0.4em\relax IEEE, 2019, pp. 496--500.

\end{thebibliography}


\begin{thebibliography}{00}
\bibitem{b1} G. Eason, B. Noble, and I. N. Sneddon, ``On certain integrals of Lipschitz-Hankel type involving products of Bessel functions,'' Phil. Trans. Roy. Soc. London, vol. A247, pp. 529--551, April 1955.
\bibitem{b2} J. Clerk Maxwell, A Treatise on Electricity and Magnetism, 3rd ed., vol. 2. Oxford: Clarendon, 1892, pp.68--73.
\bibitem{b3} I. S. Jacobs and C. P. Bean, ``Fine particles, thin films and exchange anisotropy,'' in Magnetism, vol. III, G. T. Rado and H. Suhl, Eds. New York: Academic, 1963, pp. 271--350.
\bibitem{b4} K. Elissa, ``Title of paper if known,'' unpublished.
\bibitem{b5} R. Nicole, ``Title of paper with only first word capitalized,'' J. Name Stand. Abbrev., in press.
\bibitem{b6} Y. Yorozu, M. Hirano, K. Oka, and Y. Tagawa, ``Electron spectroscopy studies on magneto-optical media and plastic substrate interface,'' IEEE Transl. J. Magn. Japan, vol. 2, pp. 740--741, August 1987 [Digests 9th Annual Conf. Magnetics Japan, p. 301, 1982].
\bibitem{b7} M. Young, The Technical Writer's Handbook. Mill Valley, CA: University Science, 1989.
\end{thebibliography}
